\documentclass[10pt,journal,compsoc]{IEEEtran}

\usepackage[pdftex]{graphicx}
\usepackage{amsmath}
\ifCLASSOPTIONcompsoc
  \usepackage[caption=false,font=footnotesize,labelfont=sf,textfont=sf]{subfig}
\else
  \usepackage[caption=false,font=footnotesize]{subfig}
\fi

\usepackage{enumitem}
\usepackage{footnote}
\usepackage[bottom]{footmisc}
\usepackage{tabularx}
\usepackage[linesnumbered,ruled]{algorithm2e}
\usepackage{multirow}
\usepackage{color}
\usepackage{url}

\usepackage{algpseudocode}
\algnewcommand{\Or}{\textbf{ or }}
\algnewcommand{\And}{\textbf{ and }}

\begin{document}

\title{Context-aware Distribution of Fog Applications Using Deep Reinforcement Learning}

\author{Nan~Wang~and~Blesson~Varghese

\IEEEcompsocitemizethanks{\IEEEcompsocthanksitem N. Wang and B. Varghese are with Queen's University Belfast, UK.\protect\\
E-mail: \{nwang03, b.varghese\}@qub.ac.uk
}}

\markboth{}%
{}


\IEEEtitleabstractindextext{%
\begin{abstract}
Fog computing is an emerging paradigm that aims to meet the increasing computation demands arising from the billions of devices connected to the Internet. Offloading services of an application from the Cloud to the edge of the network can improve the overall Quality-of-Service (QoS) of the application since it can process data closer to user devices. Diverse Fog nodes ranging from Wi-Fi routers to mini-clouds with varying resource capabilities makes it challenging to determine which services of an application need to be offloaded. In this paper, a context-aware mechanism for distributing applications across the Cloud and the Fog is proposed. The mechanism dynamically generates (re)deployment plans for the application to maximise the performance efficiency of the application by taking the QoS and running costs into account. The mechanism relies on deep Q-networks to generate a distribution plan without prior knowledge of the available resources on the Fog node, the network condition and the application. The feasibility of the proposed context-aware distribution mechanism is demonstrated on two use-cases, namely a face detection application and a location-based mobile game. The benefits are increased utility of dynamic distribution in both use cases, when compared to a static distribution approach used in existing research.
\end{abstract}

\begin{IEEEkeywords}
Fog computing, decentralised cloud, Edge computing, context-aware distribution
\end{IEEEkeywords}}

\maketitle

\IEEEdisplaynontitleabstractindextext

\IEEEraisesectionheading{
\section{Introduction}
\label{sec:introduction}}
The next generation of Cloud applications is anticipated to leverage computing resources that will be available at the edge of the network~\cite{nextgencloudcomputing-00, satya-01}. Typically, resources at the edge of the network will be more constrained in terms of processing capabilities when compared to the Cloud~\cite{fogresourcemenagement-1, cloudfuturology-1}.  
The computing paradigm that makes use of computing resources both in the Cloud and along the Cloud-Edge continuum is referred to as Fog computing~\cite{fogcomputing-00}. 

In traditional Cloud computing, applications that service end-user devices reside in Cloud data centres. Fog applications, on the other hand, will be serviced by both distant Clouds and Fog resources that are near user devices. This is done to bring latency sensitive computing to the Fog resource and make the application more responsive for the end-user~\cite{satya-01, enorm, chen2017empirical}. Similarly, the volume of data that is transferred to the Cloud for processing, which can be expensive in monetary terms, can be reduced if processed nearer to the source on Edge resources~\cite{edgecomputing-01, fogresourcemenagement-1}. The benefits are improving the overall Quality-of-Service (QoS) of the application that directly impacts a user's experience as well as making them cost-efficient

\begin{figure}
\begin{center}
	\includegraphics[width=0.5\textwidth]
	{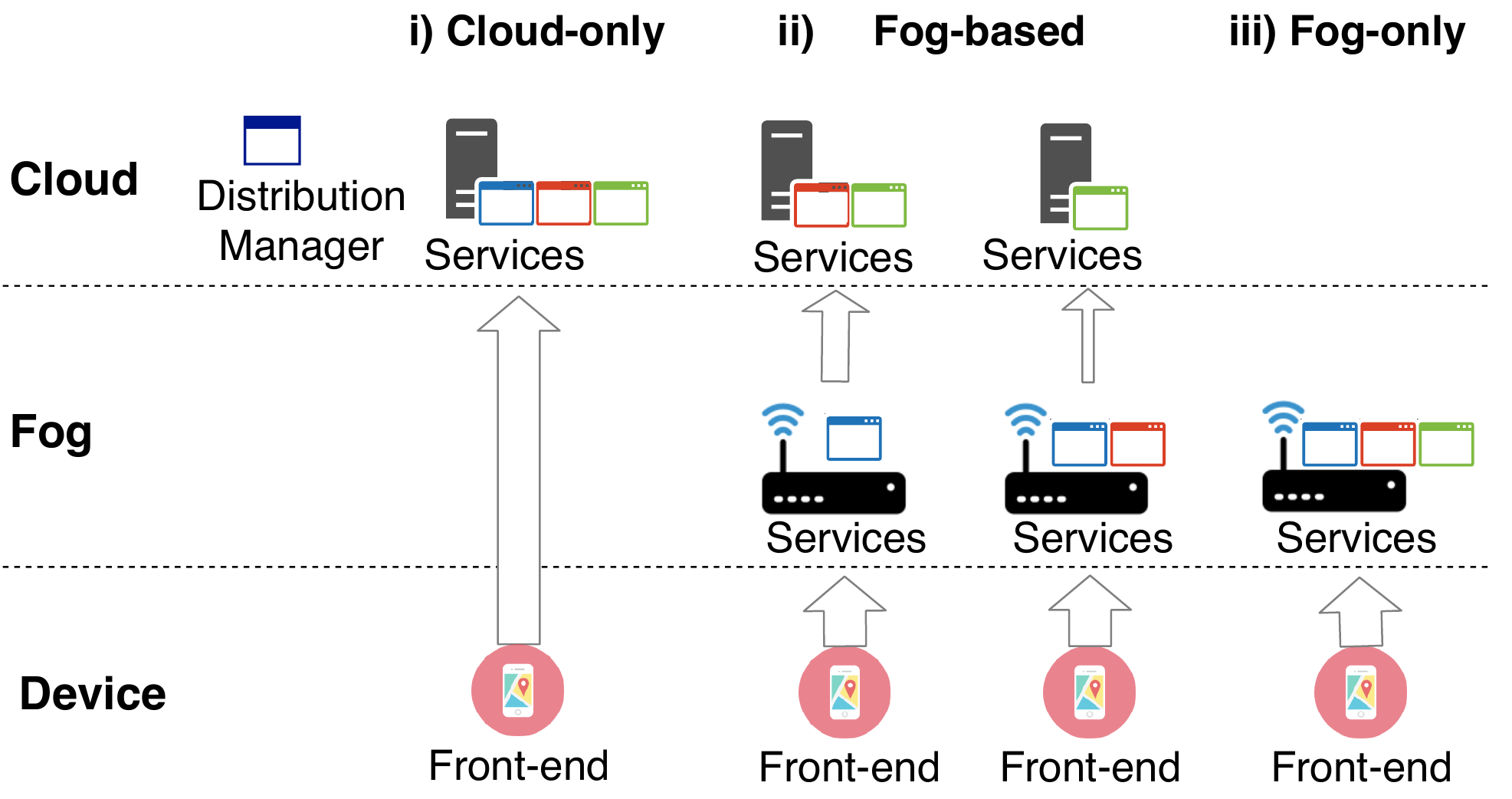}
\end{center}
\caption{Illustration of the distribution of a modular Fog application. Data movement is indicated by directed arrows, width of which represents the amount of data transmitted.}
\label{fig:FogApp}
\end{figure}

Distribution across the Cloud and the Fog can be achieved if applications are designed as a composition of multiple services~\cite{defog-01, fogservices-1, fogservices-2, fogservices-3}. To achieve a specific service level objective for an application, such as communication latency for individual user requests should not exceed a certain threshold, then certain services will need to be moved to the Fog from the Cloud. This is illustrated in Figure~\ref{fig:FogApp} in which either one or more services from the Cloud can be brought to the edge of the network, which may be geographically closer to end-users. This would require the deployment of the appropriate services in three ways: (i) Cloud-only -- all the services are hosted in the Cloud VM; (ii) Fog-based -- some of the services are hosted in a Fog node while the rest in the Cloud VM; (iii) Fog-only -- all services are hosted in a Fog node~\cite{defog-01}.

The number of services that will move to the Fog
from the Cloud will be based on the availability of Fog resources and network conditions at any given time. This becomes a complex task because the Fog is a transient environment - resource availability and utilisation and network conditions change over time. It is challenging to determine how many services need to be moved to the Fog in the face of variable system and network conditions. 

A key question that needs to be addressed in the above context given the transient nature of the Fog is \textit{which of the services and how many services of an application should be distributed across the Cloud and Fog?} Ad hoc distribution across the Cloud and Fog can be detrimental to the performance of the application and in turn, affects the QoS. Therefore, to maximise the performance of an application it is necessary to determine the best combination of services that are distributed across the Cloud and the Fog. A variety of factors will need to be considered for determining this, including resource availability and utilisation on the Fog and network conditions. Also, the economic model used at the Fog (whether the Fog is more expensive than the Cloud) is important in arriving at a decision. This naturally results in a trade-off between the overall QoS and the monetary cost. 

The research problem targeted in this article is '\textit{the distribution of Fog applications across a multi-layer Fog computing system}'. The three questions addressed to tackle the problem are: 
\begin{itemize}
    \item [\textbf{Q1}: ] When should a Cloud application make use of the Fog for improving performance? 
    \item[\textbf{Q2}: ] How many services of an application should be placed on the Fog?
    \item[\textbf{Q3}: ] What is the trade-off between QoS and monetary costs when using Fog computing?
\end{itemize}

We propose a deep reinforcement learning based approach to address the research problem. Reinforcement learning is considered a good solution for this because:
(i) it provides the capability to learn an optimal solution; (ii) it does not require a pre-trained model in contrast to other machine learning mechanisms. This means that no prior knowledge of the Fog nodes is required. In this research, Deep Q Network (DQN) are employed, which works well with continuous input and discrete output, to develop the context-aware distribution mechanism in a Fog system.

The contributions of this research are as follows:
\begin{enumerate}
    \item \textit{The formulation of the problem and a mathematical model to capture the distribution of an application across Cloud and Fog resources both in terms of the QoS of the application and the running costs on Cloud and Fog resources}. The distribution problem is long-term, and the goal is to maximise the overall benefits of a series of deployments of applications, which has not been considered in existing works.
    \item \textit{The design and development of a context-aware mechanism to distribute modular applications in a three-tiered distributed system hierarchy}. Existing research for offloading modules from the Cloud to the Fog is based on static distribution or scheduling of the application by relying on initial deployment. The mechanism we propose on the other hand distributes the modular components across Cloud and Fog resources in a dynamic manner - the decision of which components needs to be distributed is made whenever it is required once an application is deployed to improve the overall performance both in terms of QoS and running costs. 
    \item \textit{The application of deep reinforcement learning for distributing modular Fog applications}. The merit of this approach is that it can select a deployment plan without prior knowledge of resource availability on the Fog computing nodes, network conditions and Fog applications. Deep reinforcement learning supports online learning, such that a pre-trained model does not need to be used. This is highly desirable in environments such as the Fog where the applications deployed and system and network conditions change rapidly.
\end{enumerate}

The feasibility of the proposed context-aware distribution mechanism is validated using a mobile game and a face detection use-cases in a Fog environment. These workloads are a natural fit for using the Fog since they are latency critical - the response time is affected by the distance between the end device and the application server. The merit of the context-aware distribution mechanism is observed in that to select an appropriate deployment plan only has a sub-second overhead. Additionally, we observe that it effectively outperforms the other static deployment approaches for both use-cases. Through the analysis of cost-efficient, QoS-aware and hybrid strategies, and the impact of Fog pricing, it is also found that the benefits of context-aware distribution mechanism are more significant when it is optimised towards QoS and Fog resources are priced less than Cloud resources.

The remainder of this paper is organised as follows. Section~\ref{sec:background} presents the Fog system and the mathematical model that underpins it, and the problem tackled in this article. Section~\ref{sec:design} presents a context-aware methodology for distributing a modular Fog application across the Cloud and the Fog using deep reinforcement learning. Section~\ref{sec:evaluation} presents two real use-cases that can benefit from Fog computing and the experimental setup. Section~\ref{sec:results} evaluates the proposed methodology and results obtained using the chosen use-cases. Section~\ref{sec:relatedwork} discusses related research and Section~\ref{sec:conclusions} concludes this article by considering future work.

\section{Background and Problem Modelling}
\label{sec:background}
This section presents modularised applications that can be distributed across the Cloud and the Fog. Subsequently, the mathematical model for context-aware distribution of the modular components of the application in a Fog-based system is considered. 

\subsection{Modular Fog Applications}
Many existing Cloud applications are service-based and modular. For example, work-flow~\cite{thai2014optimal} and Bag-of-Task~\cite{thai2018survey} applications comprise services that are geo-distributed across clusters of systems. In a conventional two-layer computing system (from Cloud to the device), the servers of the applications run in the Cloud and the front-end on devices, such as smartphones. Modular applications are a good design fit for Fog computing in the following two ways. Firstly, services can be moved from the Cloud to the Fog so that data is processed closer to end users instead of in distant data centres to reduce latency. This is possible because the application is inherently composed of a collection of services. Secondly, modular applications lend themselves for flexible usage on Fog nodes. Since the availability of Fog resources are expected to change over time~\cite{enorm}, it may be easier to move services between the Cloud and the Fog node if they are modular. 

Currently, Cloud applications are distributed horizontally across clustered systems; resources required to meet the demands of the workload are obtained from the consolidated resources in a data centre. A different requirement in a Fog system is that the application will need to be distributed vertically across multiple tiers. Figure~\ref{fig:FogApp} compares different types of distributions of a modularised Cloud application in a three-tier Fog system. The modularised Cloud application is either the original modular applications that consist of a collection of modules or applications that are partitioned to suit a Fog system. These modules (the blue, red and green modules in Figure~\ref{fig:FogApp}) work together to provide the overall functionality of the application. 

Three distribution scenarios can be applied to a modularised Cloud application, namely the Cloud-only, the Fog-based and the Fog-only distribution. In the Cloud-only scenario, the server of the Cloud application is hosted in a Cloud VM and the front-end application is installed on an end device (e.g. smartphone). While the front-end application is active, the end device sends data to the server in the Cloud as indicated by the directed arrow in Figure~\ref{fig:FogApp}. The width of the arrow is representative of the amount of data sent.

In the Fog-based distribution scenarios, there could be several distribution plans depending on the number of modules to be deployed on a Fog node.
By applying the Fog-based distribution methods, the amount of data sent from the Fog node to the Cloud VM is expected to be reduced (arrow with reduced width) since it processed on Fog nodes and thereby reduces application response time. More modules could be deployed in the Fog node depending on the availability of resources to reduce further the amount of data transferred beyond the Fog node to the Cloud VM. In the Fog-only distribution scenario, all the modules are deployed in a Fog node such that all data received from an end device is processed in the Fog node.

Although employing Fog computing services has been proven to be effective to improve the Quality-of-Service (QoS) of Cloud applications\cite{enorm, sarkar2018assessment}, using Fog computing services will be challenging. This is because the Fog is restricted due to limited hardware resources and the availability of resources change over time due to varying network and system conditions. The question of when and how to utilise a Fog node for elastic Fog applications is not well addressed in literature, especially for applications with many modules since numerous distribution plans could be a possibility. Therefore, in this paper, the impact of different distribution plans on the performance of elastic Fog applications is explored and a context-aware distribution mechanism that generates an optimal distribution plan is proposed.

\subsection{Problem Model}
\label{sec:model}

Table~\ref{tbl:systemNotation} shows the mathematical notation employed in this research for a three-tier (Device-Fog-Cloud) computing system. The three-tiered system is referred to as the Fog computing system in this paper.
It is assumed that an application server $a$ is modular and comprises $N$ different modules. $a$ is originally hosted on a Cloud VM and the users of $a$ install the client-side application on their devices. The conventional two-tier (Device-Cloud) computing system is a typical model for delivering application services. However, in the Fog computing system, when a Fog node is available to provide computing services, a Distribution Manager of $a$ would need to decide whether a redistribution of $a$ across the new computing system is beneficial - a performance gain with reduced application latency can be obtained. The redistribution decision is to find the optimal $k \in \{0 \dots N\}$ such that the first $k$ modules of $a$ will be deployed on a Fog node that is located between the device and Cloud layers. Note that when $k=0$, $a$ is hosted in the Cloud VM as is in the original two-tier (Device-Cloud) system; when $k=N$, $a$ is hosted only on the Fog node; and when $k$ is between 0 and $N$, $a$ is distributed across the three-tier system, resulting in one part of $a$ hosted in the Fog node while the remaining part in the Cloud VM.

\begin{table}
\centering
\caption{Notation of parameters to model a Fog computing system}
\begin{tabular}{c  p{6cm}} \hline
\textbf{Parameter} & \textbf{Description} \\ \hline
\(a\) & A server of a modular application\\ 
\(k\) & The number of modules to be deployed on a Fog node in one deployment\\
$S$ & A distribution strategy containing $m$ deployment plans\\
$U$ & Utility of a Fog application \\
$m$ & Number of deployments in a distribution problem of a Fog application \\
$T$ & Completion time of one deployment \\
$R$ & Number of user requests being processed in one deployment \\
$C$ & Overall cost of one deployment\\
$C_C$ & Cost of using Cloud computing services \\
$P_C$ & Price of a reserved Cloud VM \\
$\lambda$ & Ratio of the unit price of Fog resource (CPU/memory/storage) to Cloud resources\\
$P_{cpu/mem/str}$ & Price of a unit of resource (CPU/memory/storage) in a customised Cloud VM\\
$R_{cpu/mem/str}$ & Average used units of resource (CPU/memory/storage) in a Fog node in one deployment \\
\hline
\end{tabular}
\label{tbl:systemNotation}
\end{table}

We assume that services available on a Fog node are available only at a premium, the availability of these resources change over time due to varying network and system conditions, and there is significant competition to use these services. These result in a dynamic environment. Fog applications, may also have a short life cycle on a Fog node due to varying computing capabilities that will be available on the node~\cite{tortonesi2018taming}. Selected services of a Cloud application are offloaded on to a Fog node based on current availability of resources on the Fog node, and network conditions. Resource availability on the Fog and network conditions will change over time. The initial deployment will, therefore, be less efficient, resulting in performance degradation given the changes in resource availability of the Fog node and network conditions. Therefore, the application will need to redeployed - a new configuration of the distribution (how many services need to be deployed on the Cloud and Fog will need to be determined.

In this context, the distribution strategy $S$ of $a$ will need to combine a series of deployment plans. This ensures that when the Fog computing environment changes, for example the availability of Fog computing resources change, the application adapts to mitigate any degradation in its overall performance.
The problem addressed in this paper is to \textit{determine and manage $m$ successive distributions of a modular Fog application in the face of variable resource availability and network conditions of a Fog system}. 
When $m=1$ the problem is to distribute an application in a single deployment.
When $m=2$ the problem, a distribution strategy for the application in two successive deployment rounds is generated. 

The objective of the research is to maximise the utility of the distribution strategy of a modular Fog application. Utility is defined in this article as the quantity that measures preferences over a set of distribution strategies. The utility function is frequently employed in Fog computing research, for example, to measure (i) revenue of Fog services~\cite{he2018multitier, zhang2017hierarchical} and (ii) performance of Fog networks~\cite{zhang2017fog}. In this paper, the utility function is defined in Equation~\ref{eq:utility}, which accounts for the trade-off between QoS of an application $a$ and the running cost given a distribution strategy $S$. 

\begin{align}
U(S, a) = \sum_{i=1}^{m} (\alpha \frac{T_i}{R_i} + \beta C_i(k_i))
\label{eq:utility}
\end{align}
In the utility function, $T$ is the time taken to complete a job in a single deployment and $R$ is the number of jobs processed in one deployment. The average time taken to service a single request represents the QoS of an application. Although Fog services may improve the QoS, additional costs over the cost of using Cloud services may be incurred, which needs to be accounted for. Therefore, the utility function takes the trade-off between QoS and the overall running cost $C$ into consideration. $\alpha$ and $\beta$ are weights that represent the relative importance assigned to the QoS and cost factors. For example, when $\beta$ is 0, the distribution strategy that leads to the highest QoS is selected, and when $\alpha$ is 0, the distribution strategy that results in the lowest running cost is considered suitable. The running cost of one deployment is a function of $k$, defined as:
\begin{align}
    C(k)= 
\begin{cases}
    C_C,& \text{if } k=0\\
    C_F, & \text{if } k=N\\
    C_C + C_F,              & \text{otherwise}
\end{cases}
\end{align}where $C_F$ and $C_C$ are the costs entailed by employing Fog and Cloud computing services, respectively. The cost of Cloud services is defined on a subscription (pay-as-you-go) basis, which is the popular pricing model adopted by public Cloud service providers such as Amazon Web Services:
\begin{align}
C_C = P_C \cdot T
\label{eq:costCloud}
\end{align}The cost of Fog computing services is defined on a fine-grained pay-for-resource-used basis:  
\begin{align}
C_F = \lambda (P_{cpu}, P_{mem}, P_{str}) \cdot (R_{cpu}, R_{mem}, R_{str}) \cdot T
\label{eq:costFog}
\end{align}
where $P_{cpu/mem/str}$ is the unit price of Cloud resources including CPU, memory and storage. 
$\lambda$ represents how expensive Fog resources are when compared to the Cloud resources. In this model $\lambda$ is chosen from $[0.001, 0.01, 0.1, 1]$, i.e. the Fog resources could be defined as 0.001, 0.01, 0.1 or 1 times the price of Cloud resources. In this paper, it is assumed that Fog resources are less expensive (or of the same price) when compared to Cloud resources. This may be an incentive for Cloud-native application providers to make use of the Fog. 

$R_{cpu/mem/str}$ is the average amount of Fog resources used in a single deployment. 

The motivation for the above Fog pricing is as follows. Fog applications may have multiple deployments to respond to the system or network changes in the environment, and in each deployment, the modules deployed on a Fog node may vary. Therefore, it is not efficient to reserve the same amount of Fog resources for an application. For example, if resources are reserved for the maximum number of modules ($k=N$), then for deployments with fewer modules moving to the Fog ($k<N$) some of the already paid for Fog resources will not be utilised, and vice-versa. Hence, a fine-grained pricing model for Fog resources is required to support each deployment of modular Fog applications.

\section{Context-aware Distribution}
\label{sec:design}
A naive mechanism for distributing a Fog application may be based on generating a static distribution strategy. For example, the same number of services are always deployed in a Fog node and the remaining services (if any) in a Cloud VM. While this strategy can be easily implemented and used in a real-time setting, it does not consider the availability of resources on a Fog node, the network conditions between the Fog node and the Cloud VM, and also the heterogeneity of Fog nodes if multiple resources are viable deployment options at the edge of the network. This results in static distribution strategies becoming quickly obsolete and applications under performing.  

A dynamic distribution strategy would overcome the above shortcomings by deploying a varying number of services in a Fog node given the context. Therefore, there is a need for a decision-making process that adapts to the context -- variable Fog nodes, Fog resource availability and network conditions. To achieve this, the utility, a measure of system performance, defined in Equation~\ref{eq:utility} for every possible deployment plan is used in this paper to compare the effectiveness of different deployment plans. By comparing the utility values of an application being deployed on the Cloud or on the Fog computing platform, it is possible to identify the best deployment plan of the application. Furthermore, if the utility value can be accurately estimated, then it is possible to proactively select the best deployment plan in advance. This result in avoiding poor application performance due to selecting an ad hoc deployment plan.

Supervised learning techniques have been frequently used in Fog computing research to estimate system performance~\cite{osanaiye2017cloud} and cost~\cite{su2018secure}. These techniques require a predictive model that will need to be trained on benchmarking data that is obtained from heterogeneous Fog nodes and all potential applications that may use Fog nodes. Such a method is feasible if the Fog nodes in the computing environment are homogeneous (same system specification) and only a few applications need to use the Fog. Supervised learning is impractical when there is a large volume of heterogeneous resources and a variety of applications with different system requirements as seen in real world computing environments. 

Reinforcement learning is an alternative to the above and is suitable to generate dynamic distribution strategies. This technique does not require pre-trained models, unlike supervised learning techniques. Using pre-trained models for predicting system performance usually requires the collection of a large amount of training data, which ideally covers many different features of the computing environment. For example, hardware specifications of a system and computation workloads deployed on the system. Therefore, creating accurate pre-trained models for different applications is cumbersome and impractical given the variety of Fog computing nodes and the number of service combinations when an application is partitioned. Therefore, reinforcement learning is considered a suitable alternative to supervised learning.
operate in an environment comprising heterogeneous Fog nodes and a variety of different Fog applications. In the following section, the distribution problem of Fog applications as a reinforcement-learning task is presented and then the context-aware distribution mechanism using Deep Q-Network (DQN) is proposed.


\subsection{Reinforcement Learning}
Reinforcement learning is defined as a process to learn best actions based on rewards or punishments. A reinforcement learning problem comprises the following six concepts: (i) agent -- the reinforcement learning algorithm; (ii) environment -- a physical world in which the agent operates; (iii) state -- the current situation of the environment; (iv) action -- the operation the agent takes; (v) reward -- the feedback from the environment based on the action the agent takes. The goal of the agent is to collect as much reward as possible through interacting with the environment by trial and error (i.e. taking different actions based on the states observed) using feedback on its actions. 

In this paper, we formulate the distribution problem by transforming the above concepts into a Fog context: (i) the agent in our problem is a Distribution Manager, who is located in a Cloud VM and responsible for distributing a modular Fog application; (ii) the environment is a Fog node, where some or all of the services of the Fog application will be deployed into; (iii) the state is the current representation of the Fog node, which accounts for the hardware specifications and resource availability of the Fog node, and the network condition between the Fog node and the Cloud VM; (iv); the action is to select a $k$ values, which is the number of services of a Fog application to deploy into the Fog node; (v) the reward is the utility defined in Equation~\ref{eq:utility}, which is a measurement of the quality of a deployment plan. 

\begin{table}
\centering
\caption{Factors used in the state vector in the reinforcement-learning task}
\begin{tabular}{c  p{6cm}} \hline
\textbf{Parameter} & \textbf{Description} \\ \hline
$CPU_u$ & Current system-wide CPU utilisation of a Fog node \\
$CPU_n$ & Number of logical CPUs in a Fog node\\
$CPU_f$ & Current CPU frequency of a Fog node\\
$MEM_p$ & Total physical memory in a Fog node\\
$MEM_{pu}$ & Current physical memory usage of a Fog node\\
$MEM_s$ & Total swap memory of a Fog node\\
$MEM_{su}$ & Current swap memory usage of a Fog node\\
$STR_d$ & Total disk space in a Fog node\\
$STR_{du}$ & Current disk usage of a Fog node\\
$IO_r$ & System-wide number of reads from the disk in a Fog node\\
$IO_w$ & System-wide number of writes to the disk in a Fog node\\
$IO_{rb}$ & System-wide number of bytes read from the disk in a Fog node\\
$IO_{wb}$ & System-wide number of bytes written to the disk in a Fog node\\
$IO_{bs}$ & System-wide number of bytes sent from a Fog node\\
$IO_{br}$ & System-wide number of bytes received by a Fog node\\
$IO_{ps}$ & System-wide number of packets sent from a Fog node\\
$IO_{pr}$ & System-wide number of packets received by a Fog node\\
$D_{FC}$ & Network delay between a Fog node and a Cloud VM\\
$D_{EC}$ & Network delay between an end device and a Cloud VM\\
\hline
\end{tabular}
\label{tbl:state}
\end{table}

In the problem defined above, the state of a Fog node at a certain time consists of 19 factors (Table~\ref{tbl:state}) related to: (i) the processor in a Fog node (i.e. $CPU_u, CPU_n, CPU_f$); (ii) the memory in a Fog node (i.e. $MEM_p, MEM_pu, MEM_s, MEM_su$); (iii) the storage in a Fog node (i.e. $STR_d, STR_{du}$); (iv) the disk I/O in a Fog node (i.e. $IO_r, IO_w, IO_{rb}, IO_{wb}$); (v) the netwrok I/O in a Fog node (i.e. $IO_{bs}, IO_{br}, IO_{ps}, IO_{pr}$); and (vi) the network conditions (i.e. $D_{FC}, D_{EC}$). These factors are selected to be representatives of the specification and computing capability of a Fog node, the network condition and the relative distance between the Fog node and the Cloud VM. Theses factors have continuous values. The set of actions is the possible values of $k$, which are $N+1$ discrete values. 

\subsection{DQN-based Distribution Mechanism}
Q-learning is a typical technique in reinforcement learning, which learns an optimal policy to maximise the total reward over several successive steps. This is suitable for the long-term distribution problem of applications defined in this paper. Due to the changing availability of Fog resources, the state of a Fog-based application is also expected to change over time. For example, moving a service to a different Fog node or changing the service that executes on the Fog. Therefore, the performance of the application may change over different redeployment rounds, which is different from the Cloud where the deployment of an application may not happen frequently over a short time period.

Q-learning is designed for a problem with discrete state and action spaces, but the state space in our problem is continuous as described above. Discretising our state-space would result in a large space and thus incurs more memory consumed by the algorithm. Therefore, we choose the DQN algorithm as an alternative to training the Distribution Manager. Deep Q-Network is a combination of deep learning and Q-learning, which uses deep neural networks to represent the mapping between states and actions in reinforcement learning~\cite{mnih2015human}.

\begin{table}
\centering
\caption{Parameter of the Deep Q Network}
\begin{tabular}{c  p{6cm}} \hline
\textbf{Parameter} & \textbf{Description} \\ \hline
$\epsilon$ & Exploration rate of the DNN in the Cloud agent\\
$\epsilon_{min}$ & Minimum exploration rate of the DNN in the Cloud agent\\
$\gamma$ & Discount rate to calculate the future discounted reward in the DNN in the Cloud agent\\
$\upsilon$ & Decay rate to decrease the number of explorations as the DNN gets good at predictions\\
$batchSize$ & Number of randomly sample experiences to replay in the DQN\\
$d$ & Maximum number of experiences to store in the replay memory $D$ in the DQN\\
\hline
\end{tabular}
\label{tbl:dqn}
\end{table}

\begin{figure}
\begin{center}
	\includegraphics[width=0.5\textwidth]
	{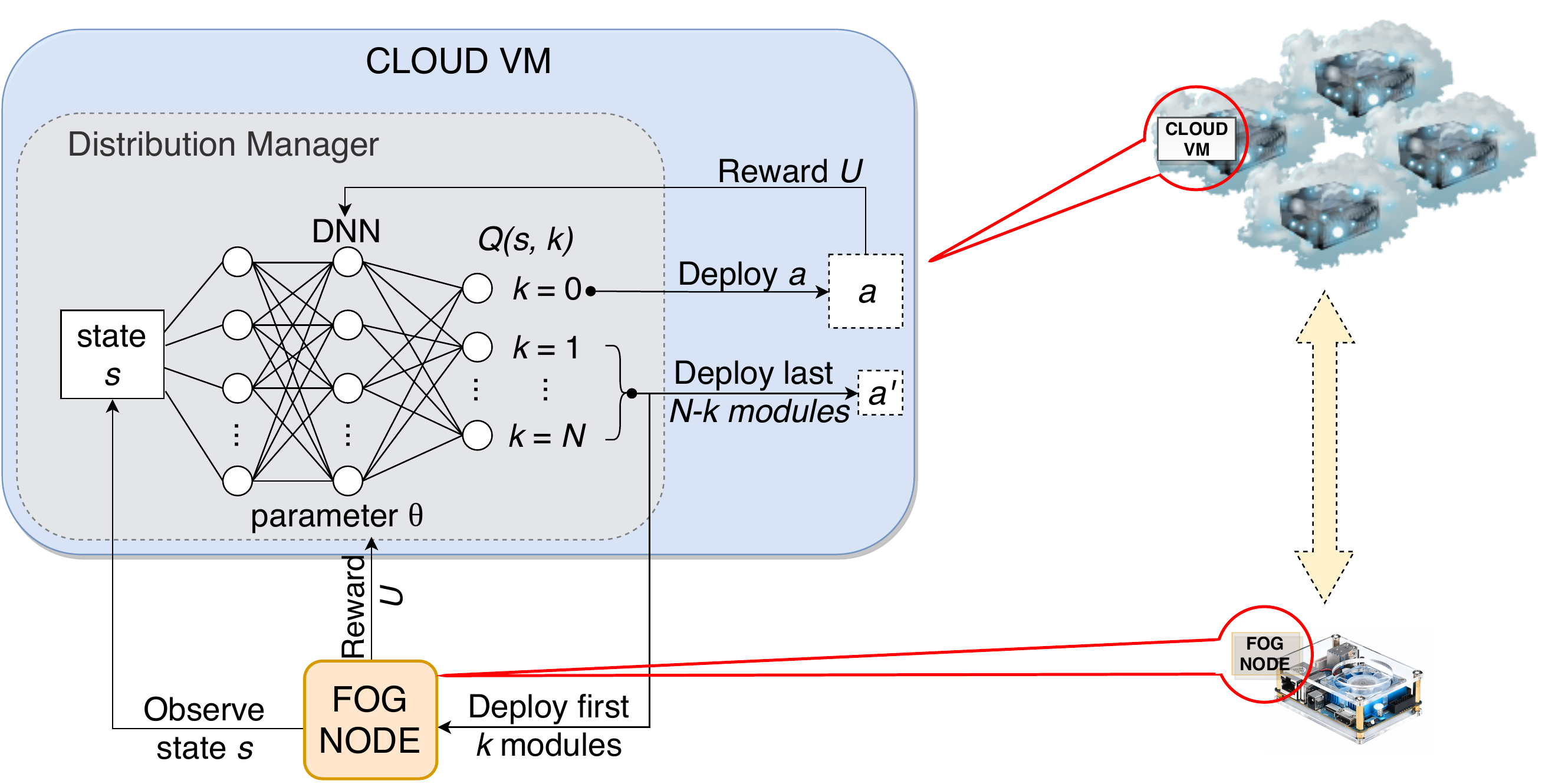}
\end{center}
\caption{A deep Q-network for a Fog computing system.}
\label{fig:dqn}
\end{figure}

Figure~\ref{fig:dqn} describes the proposed context-aware distribution mechanism using DQN in a Fog environment. It involves a Distribution Manager, a set of state, a set of actions per state (i.e. $k$ values). The state $s$ of a Fog node is requested before each deployment is made by the Distribution Manager in a Cloud VM. When a Fog application is to be distributed, the current state is fed into a deep neural network. Different from the traditional Q-learning algorithm, which requires a look-up table for pairing states and actions, the deep neural network adopted in the deep Q-network algorithm is a better fit for our problem as the continuous state factors would result in a huge state space without an approximation function.

The state of the Fog computing environment needs to be known for deploying $k$ modules to the Fog node and $N-k$ modules on the Cloud VM. 
Selecting a particular $k$ value for the distribution plan relative to a specific state provides the Distribution Manager with a reward (i.e. the utility of a single deployment). The goal of the Distribution Manager is to maximise the accumulated utility over $m$ episodes of deployment. It does this by adding the maximum utility attainable from future states to the utility for achieving its current state, effectively influencing the current selection of $k$ value by the potential future utility. This potential utility is a weighted sum of the expected values of the utility of all future steps starting from the current state.

The deep neural network approximates a Q function between the states and $k$ values, and updates the model during a series of deployments. Initially, the Distribution requests an observed state from a Fog node and randomly deploys the first $k$ modules on the node. After this deployment has been completed (i.e. all user requests have been processed), the job completion time $T$ and used Fog resources $R_{cpu/mem/str}$ are reported to calculate the utility of this deployment. Subsequently, the Distribution Manager uses the utility as feedback to the neural network. In the following deployments, the Distribution Manager starts to learn a predictive model to find the optimal $k$ from the $N+1$ distribution plans such that the overall utility of $m$ successive deployments is maximised.  

\begin{algorithm}
\DontPrintSemicolon
Initialise replay memory $D$ with capacity $d$\;
Initialise k-value function $Q$ with random weights $\theta$\;
Initialise target k-value function $\hat Q$ with weights $\hat \theta$\;
\ForEach{episode}{
Request the Fog node for the initial state $s_1$\;
  \For{$j \in \{1,\dots,m\}$}{
    With probability $\epsilon$ select a random value $k$, otherwise select $k_j = argmax_k Q(s_j, k; \theta)$\;
Deploy the fist $k$ modules of application $a$ in the Fog node\;
Deploy the rest modules of $a$ in the Cloud VM\;
    Observe $T, R_{cpu}, R_{mem}, R_{str}, s_{j+1}$ and calculate utility $U_j$\;
    Set $s_{j+1} = s_j$\;
    Store $(s_j, k_j, U_j, s_{j+1})$ in replay memory $D$\;
        \If{$len(D) > batchSize$}{
Sample random minibatch of $batchSize$ from $D$\;
\ForEach{$(s_q, k_q, U_q, s_{q+1})$}{
    \uIf{episode terminates at step $q + 1$}{$y_q =U_q$}
    \Else{
    $y_q =U_q + \gamma max_{k'} \hat Q(s_{q+1}, a';\hat \theta)$}
Perform a gradient descent step on $(y_q - Q(s_q, k_q;\theta))^2$ with respect to $\theta$\;
$Q = \hat Q$\;}
\If{$\epsilon > \epsilon_{min}$}{
$\epsilon = \upsilon \epsilon$\;
}
}
  }
}
\caption{DQN-based distribution mechanism}
 \label{algo:dqn}
\end{algorithm}

Algorithm~\ref{algo:dqn} describes the DQN-based distribution mechanism of a Fog application. The parameters of the DQN is listed in Table~\ref{tbl:dqn}. The neural network takes an input state $s$ described above and outputs $Q$ values over all possible distribution plans (i.e. $k$ values). Episodic training is considered for the scenario of $m$ successive deployment episodes. The DQN first initialises a replay memory specified capacity (Line~1). This is the experience replay mechanism in a DQN that reuses previous experience to improve model performance. Then the neural networks for both the $Q$ function and the target $\hat Q$ are initialised with random weights~(Lines~2--3). In each episode, the Fog node is requested for the initial state~(Line~5), and then $m$ deployment requests are processed~(Line~6). The episode terminates when the $m^{th}$ deployment is completed. In each deployment, $k$ is selected either randomly with exploration rate $\epsilon$ or by the current $Q$ function~(Line~7). Consequently, the Distribution Manager deploys the associated modules on the Fog node and the Cloud VM, respectively~(Lines~8--9). After the jobs in this deployment are completed, the Distribution Manager receives the job completion time and Fog resources usage to calculate the utility of this deployment. A replay memory is implemented so that past experiences (including $s$, $k$, $U$) is remembered and can be reused to train the model efficiently. The next state of the Fog node is also remembered~(Lines~10--12). To learn from past experiences, the Distribution Manager takes a random sample from its replay memory and train the current model to minimise the loss, which is the squared difference between the target and the predicted $Q$ values~(Lines~13--22). The exploration rate $\epsilon$ is continuously reduced as the model performance gets better~(Lines~24--26). By defining a minimum exploration rate $\epsilon_{min}$ it is ensured that the Distribution Manager explores for at least this amount of time. 

\textbf{Strategies}: through tuning $\alpha$ and $\beta$, we are able to define different strategies of the context-aware distribution approach, including
\begin{itemize}
    \item \textit{Cost-effective}: when $\alpha$ is 0, the Distribution Manager takes a pure cost-effective approach by training the DQN to minimise the overall cost of redistribution. 
    \item \textit{QoS-aware}: when $\beta$ is 0, the Distribution Manager takes a pure QoS-aware approach by training the deep Q-network to minimise the average application latency of the redistribution. 
    \item \textit{Hybrid}: when $\alpha$ and $\beta$ are not 0, the Distribution Manager takes a hybrid approach and considers the trade-off between QoS and cost by training the deep Q-network to minimise the weighted sum of these two factors.
\end{itemize}
 As comparisons to the context-aware approaches with above strategies, $N+1$ static distribution approaches of a Fog application are considered. When employing a static distribution approach, the Distribution Manager keeps deploying the same $k$ modules of $a$ on a Fog node in the $m$-episode distribution. The static distribution approaches do not consider the computing capability, resource availability, the location and network condition of a Fog node.

\section{Experimental Evaluation}
\label{sec:evaluation}

In this section, the context-aware distribution approaches presented in Section~\ref{sec:design} are evaluated. The experimental setup, including the Fog use-cases and the hardware platform employed in this research, is firstly presented. This is followed by evaluating the performance of the context-aware distribution mechanism against a variety of metrics, including system overhead and utility as defined in Equation~\ref{eq:utility}. The evaluation considers different values of the parameters used in Equation~\ref{eq:utility} and Equation~\ref{eq:costFog}.

\subsection{Fog Use-Cases}
Two applications are employed for evaluating the context-aware distribution mechanism proposed in this paper. The first is a real-time face detection application, and the second is a location-based mobile game.
Both use-cases are server-based and a natural fit for Fog computing since they are latency critical~-- response time is affected by the distance between user devices and the server. Hence, a subset of the functionality of the Cloud server can be brought closer to devices on the Fog node.

The chosen use-cases are also representative of different workloads that can benefit from Fog computing: the mobile game represents an online application whose Fog server responds to incoming user requests; the face detection workload, in contrast, is representative of a data-intensive streaming application, in which case the Fog server pre-processes incoming data and relays it to the Cloud.

\subsubsection{Real-time Face Detection}
The original face detection application is Cloud server-based such that an end device with an embedded video camera captures a continuous video stream and transmits it to the Cloud server. The goal of the application is to detect faces from each individual video frame by using Pillow\footnote{\url{https://pillow.readthedocs.io}} and OpenCV\footnote{\url{https://opencv.org}}. 

\begin{figure}
\begin{center}
	\includegraphics[width=0.5\textwidth]
	{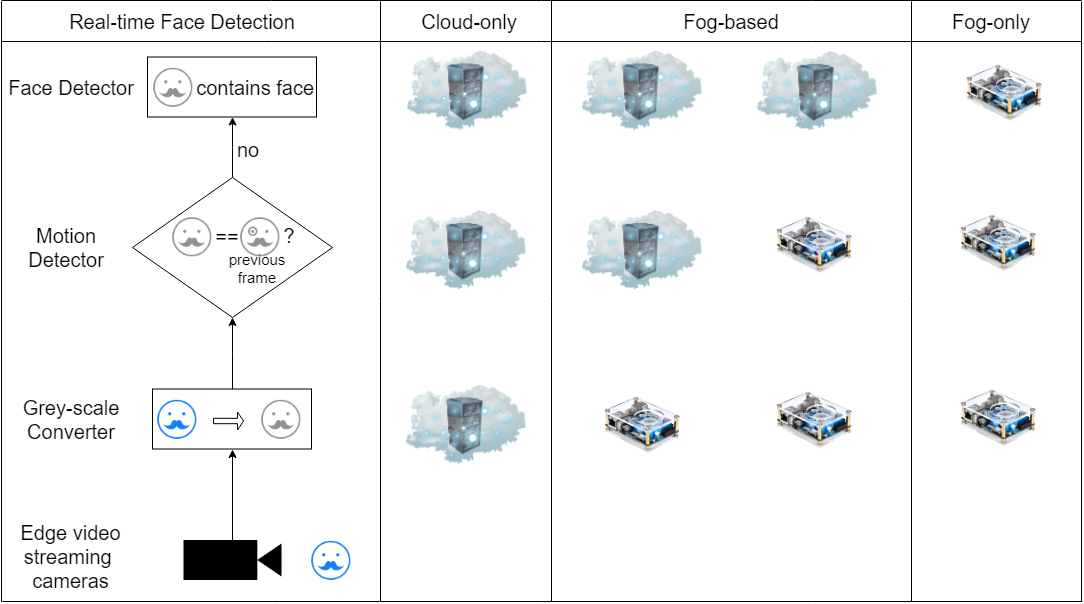}
\end{center}
\caption{The Cloud-only, Fog-based and Fog-only distribution options for the real-time face detection use-case.
}
\label{fig:realtimefacdetection}
\end{figure}

Typically, the application streams the video to the Cloud, and all processing is performed on the Cloud server. By employing Fog computing (data processing near where it is generated), the amount of data transferred to the Cloud can be reduced, thereby minimising communication latencies, while obtaining reasonable overall system performance. The application is a good fit for Fog computing since firstly, it is latency-critical and bandwidth consuming~-- response time is heavily affected by the distance between user devices and the Cloud server. Secondly, a subset of the services from the Cloud can be brought closer to devices to reduce the amount of data transferred to the Cloud. In the experiments, the images captured in real-time are recorded once and repeatedly used (the same workload applied to different approaches) in order to ensure that our comparisons are even. 

The server application is modular and comprises the following three services ($N = 3$) for detecting faces from a frame of the video as shown in Figure~\ref{fig:realtimefacdetection}: (i) \textbf{\textit{Grey-scale converter}} is a data pre-processing service, which is used by many object detection algorithms for reducing the amount of computation done on an image that would otherwise be required if it was a colour image. As a result of this component, only one-third the size of data will be processed in the remaining procedures. (ii) \textbf{\textit{Motion detector}} is a data filtering service. Consider, for example, the use case of real-time face detection used in security surveillance using CCTV cameras, which would normally send multiple video frames in one second. In these use-cases, a condition check on successive video frames would effectively reduce computations spent on similar frames (consider for example a security feed in a static environment, such as a home or museum). (iii) \textbf{\textit{Face detector}} is a computationally expensive service that identifies frontal human faces in a video frame using machine learning.

This application can be distributed in the following four ways as shown in Figure~\ref{fig:realtimefacdetection}: (i) Cloud-only services($k=0$) -- all the services mentioned above are deployed in a Cloud VM; (ii) Fog-based pre-processing ($k=1$) -- the grey-scale converter is deployed in a Fog node and the other services deployed in a Cloud VM; (iii) Fog-based data filtering ($k=2$) -- the grey-scale converter and the motion detector are deployed in a Fog node and  the face detector is deployed in a Cloud VM; (iv) Fog-only services ($k=3$) -- all the services are deployed in a Fog node. If the same distribution plan from the above four is employed for a series of deployments, then the approach is considered as static. The context-aware distribution mechanism proposed in this paper dynamically (re)selects one of the above four distribution plans in order to maximise performance by considering the varying states of a Fog node presented in Section~\ref{sec:design}.

\begin{table}
\centering
\caption{Default parameter values}
\begin{tabular}{c  p{1cm}| c p{1cm} } \hline
\textbf{Parameter} & \textbf{Value} & \textbf{Parameter} & \textbf{Value} \\ \hline
$\lambda$ & 0.01 & $\gamma$ & 0.95 \\
$P_{str}$ & 0.000032 & $batchSize$ & 5 \\
$m$ & 20 & $\epsilon$ & 1 \\
$P_{mem}$ & 0.005458  & $hiddenLayer$ & 2\\
$\alpha$ & -1 & $\epsilon_{min}$ & 0.01 \\
$P_{cpu}$ & 0.04073 & $learningRate$ & 0.001\\
$\beta$ & -1 & $\upsilon$ & 0.99 \\
$hiddenLayerNode$ & 24 & $P_C$ & 0.0132  \\
\hline
\end{tabular}
\label{tbl:defaultParam}
\end{table}

\subsubsection{Location-based Mobile Game}

The application is an open-source mobile game similar to Pok\'emon Go, named iPokeMon. iPokeMon comprises a client\footnote{\url{https://github.com/Kjuly/iPokeMon}} for the iOS platform, which can be used on mobile devices, and a server\footnote{\url{https://github.com/Kjuly/iPokeMon-Server}} that is hosted on a public Cloud. User navigates through an environment in which virtual creatures named Pok\'emons are distributed. 
The iPokeMon game server was redesigned to be hosted on the Cloud and a Fog node. The Fog hosts a game server that handles requests from recognised users whose data exists in the game database; for example, to update the players' tamed Pok\'emons. The Cloud hosts the original iPokeMon server that is able to handle requests from new users whose data does not exist in the game database; for example, to create a user with default Pok\'emons profiles. The local view on the Fog server is updated by frequent requests sent to the Fog server. If user requests are serviced from a distant data centre, then user experience is affected due to lags in refreshing. Hence, the Fog is beneficial to reduce latency for this workload. 

The iPokeMon server is tested using Apache JMeter\footnote{\url{http://jmeter.apache.org/}}. 100 HTTP requests during a connection (a user is playing the iPokeMon game) between the user device and the original Cloud server is recorded. During this time, the number and type of requests and the parameters sent through the requests are recorded. Subsequently, JMeter replays the user requests in the experiments.

\begin{figure}
\begin{center}
	\includegraphics[width=0.5\textwidth]
	{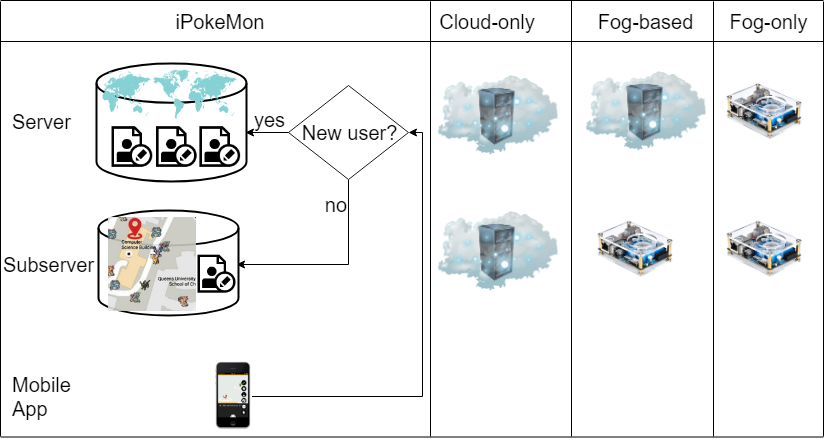}
\end{center}
\caption{The Cloud-only, Fog-based and Fog-only distribution options for the iPokeMon use-case.
}
\label{fig:ipokemon}
\end{figure}

This application could be distributed in three ways as illustrated in Figure~\ref{fig:ipokemon}: (i) Cloud-only services ($k=0$) -- the original iPokeMon server is hosted in a Cloud VM; (ii) Fog-based services ($k=1$) -- the functionality and database to service existing users are hosted in a Fog node and the functionality and database to service new users are hosted in a Cloud VM; (iii) Fog-only services ($k=2$) -- the original iPokeMon server is hosted in a Fog node. The static and context-aware approaches for distributing iPokeMon is defined the same as for FD above.

\subsection{Experimental Setup and Approaches Evaluated}

\textit{\textbf{Setup:}} A Device-Fog-Cloud system is setup using a laptop connected to a router via Wi-Fi, an ODROID-XU4 board connected to the same router via Ethernet, and a t2.micro VM which is running Ubuntu 14.04 LTS provided by Amazon Web Services Elastic Compute Cloud from its Dublin data centre. The Fog node is located in the Computer Science Building of Queen's University Belfast in Northern Ireland. The board has 2~GB of DRAM memory, and one ARM Big.LITTLE architecture Exynos 5 Octa processor running Ubuntu 14.04 LTS. 

Table~\ref{tbl:defaultParam} displays the default values of parameters used in the experiments.
As defined in Section~\ref{sec:model}, the distribution problem is considered as a series of $m$ successive deployments and the objective is to maximise the overall utility defined in Equation~\ref{eq:utility}. A total of 20 deployments 
are considered in each experiment for the Distribution Manager to learn for 1,000 episodes, which results in the experiment being run for over 33 hours. 
The prices of Cloud computing resources ($P_C, P_{cpu/mem/str}$) is defined as US dollars per hour for a single unit of the resources (a VM, vCPU core, 1-GB memory, 1-GB storage). This is how current Cloud Infrastructure-as-a-Service such as Google Compute Engine\footnote{\url{https://cloud.google.com/compute/pricing}} is priced. In the context-aware distribution mechanism, a DQN with two hidden layers and 24 nodes is learnt during the 20 deployments~\ref{algo:dqn} to model the relationship between the system state of a Fog node and the $k$ value. The DQN is implemented with the Keras\footnote{\url{https://keras.io/}} deep learning library, with the learning rate of 0.001 and 5 memorised experience to replay.

To create a realistic Fog computing node, which is likely to have variable resource availability over time, the Linux package \textit{stress}\footnote{\url{http://manpages.ubuntu.com/manpages/trusty/man1/stress.1.html}} is chosen to stress test the Fog node. The CPU and memory of the Fog node are divided into eight units, each unit comprising 1 CPU core and 256 megabytes of memory. For the experiments in Section~\ref{sec:performance} and Section~\ref{sec:parameter},  the Fog node was stressed continuously - the cores executes a workload that consumes a random number between 0 and 7 and the memory is flooded using the stress package. The random number is changed every 10 seconds. This ensures that the system state changes on the Fog node. The same sequence of the stress tests is applied to all distribution approaches to ensure that system states are similar in all experiments.

\textit{\textbf{Approaches Evaluated}}:
Since the application relies on three services, there are consequently four static approaches (the Cloud-only services, Fog-based pre-processing, Fog-based data filtering and the Fog-only services) with $k \in \{0 \dots 3\}$. A static approach (\textit{S-k}) means that for the 20 successive deployments in each distribution, no matter what the state of the Fog node is, the Distribution Manager always deploys the first $k$ modules on the Fog node and the remaining $N-k$ modules on the Cloud VM. On the contrary, the context-aware approaches (\textit{Context-aware}) employ Algorithm~\ref{algo:dqn} and dynamically assign a $k$ value using DQN in each deployment.

\section{Results}
\label{sec:results}
The experiments provide insight into the benefits of using the context-aware distribution mechanism of elastic Fog applications. The system overhead of the online decision-making process is discussed before comparing the three context-aware approaches against four static distribution approaches. A further discussion on the impacts of varying parameters in the system is presented.

\subsection{Training and Overhead}

\begin{figure}
\begin{center}
	\subfloat[FD]
	{\label{fig:FDtrain}
	\includegraphics[width=0.5\textwidth]
	{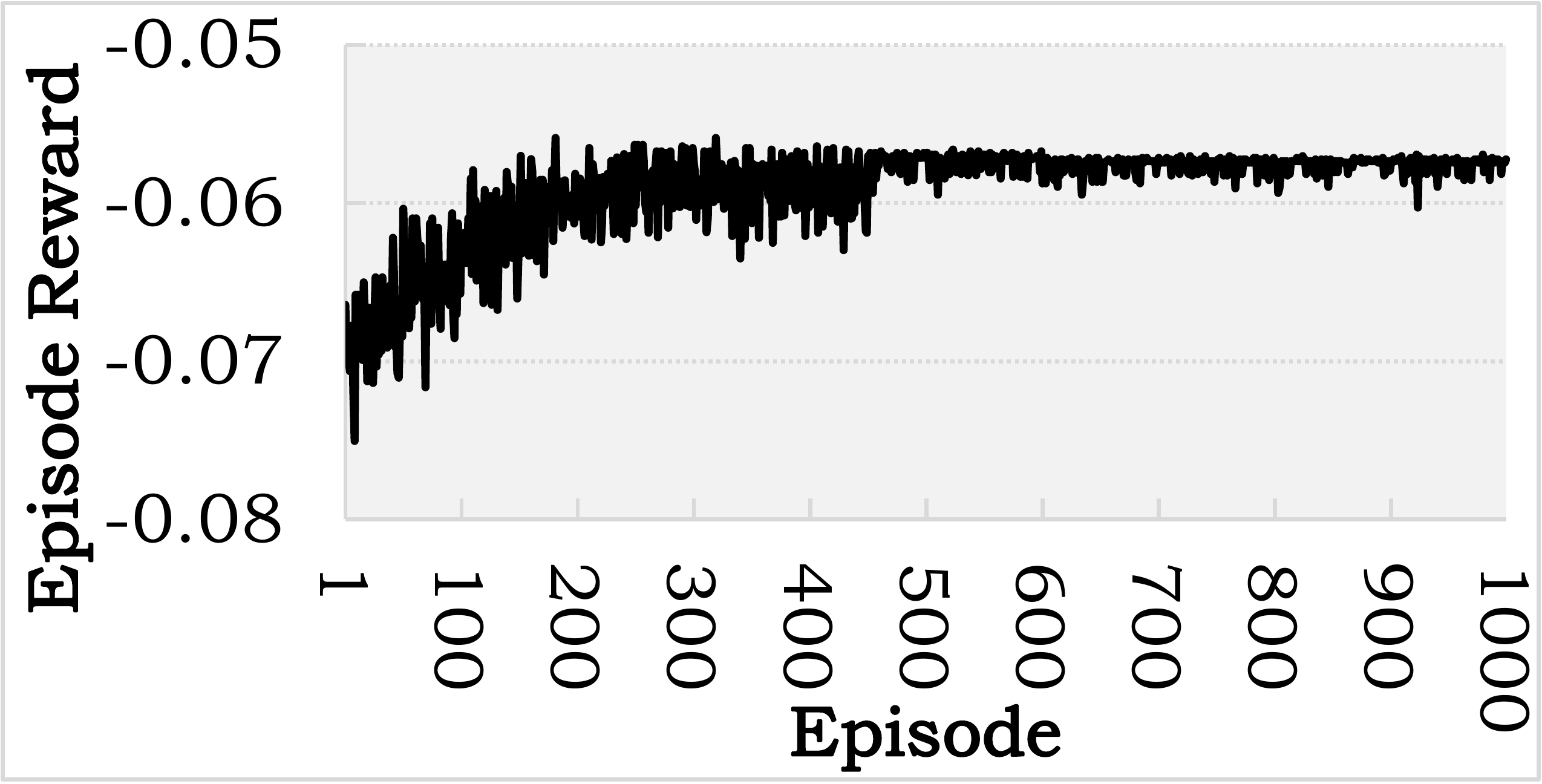}}
	\hfill
	\subfloat[iPokeMon]
	{\label{fig:PMtrain}
	\includegraphics[width=0.5\textwidth]
	{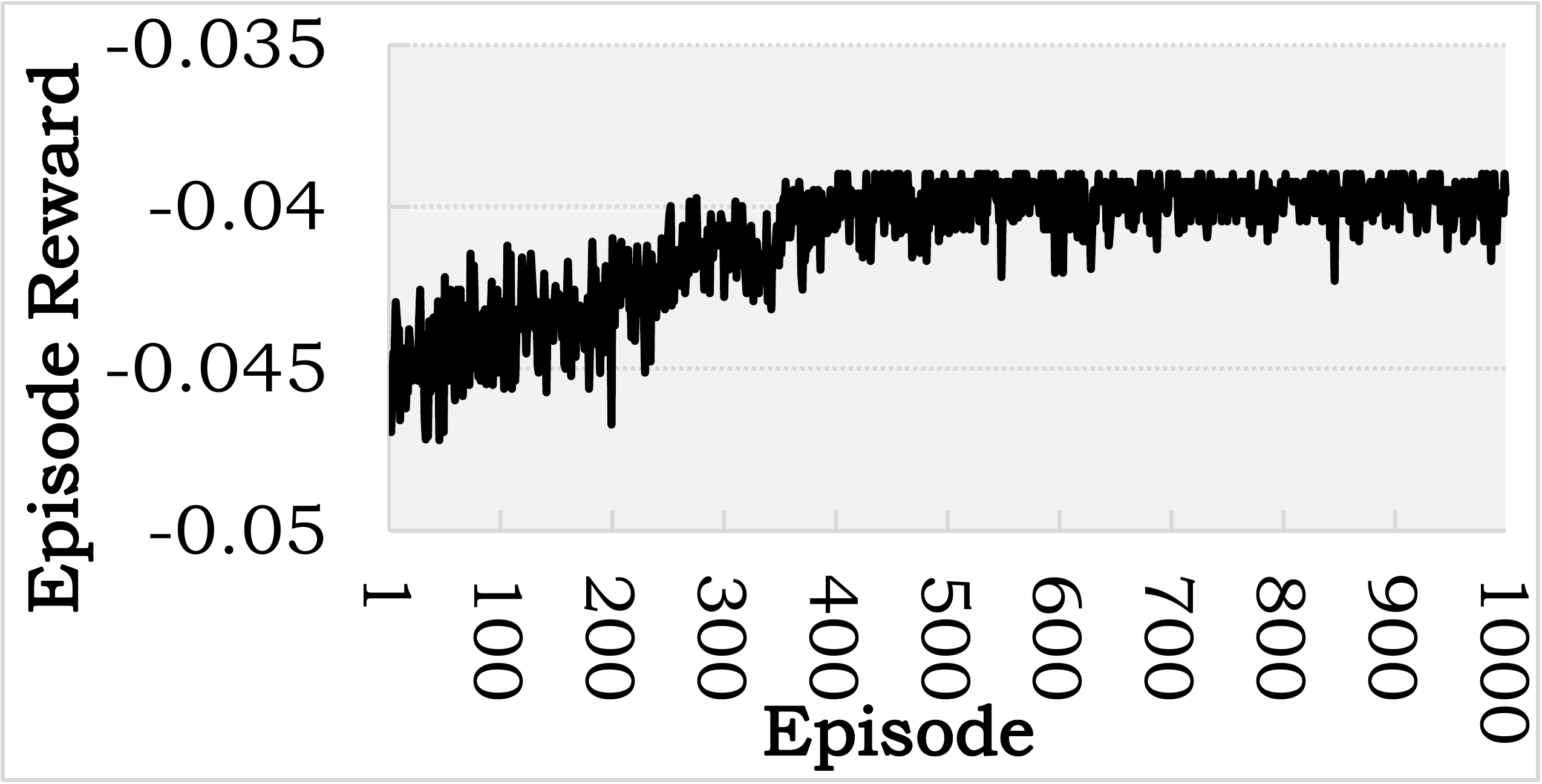}}
\end{center}
\caption{Learning curve of the hybrid context-aware approach showing the training improves the episode reward.}
\label{fig:training}
\end{figure}

Figure~\ref{fig:training} displays the episode reward (i.e. the accumulated utility of 20 successive deployments) when the DQNs in the Distribution Manager of the FD application\ref{fig:FDtrain} and iPokeMon\ref{fig:PMtrain} for up to 1,000 episodes. For both use-cases, the episode reward is observed to gradually increase with the number of episodes, which means the DQN is getting better performance with more experiences. The episode reward reaches and stays in its maximum point from around 600 and 400 episodes, at which point a total of 12,000 and 8,000 deployments have been carried out by the Distribution Manager for FD and iPokeMon respectively. This means the DQNs have learnt an optimal distribution plan for the 20 successive deployments and could not increase the utility any more. Therefore, for the analysis of the context-aware distribution approaches (including QoS-aware, cost-effective and hybrid) in the following sections, we present the DQNs trained with 600 and 400 episodes for FD and iPokeMon respectively, except specified otherwise.

\begin{table}
\centering
\caption{Statistics of the time (in milliseconds) taken by each decision making of the hybrid DQNs trained for FD and iPokeMon.}
\begin{tabular}{c c c c c c} \hline
\textbf{Min.} & \textbf{1st Quartile} & \textbf{Median} & \textbf{Mean} & \textbf{3rd Quartile} & \textbf{Max.}\\ \hline
85 & 124 & 136 & 139.5 & 149 & 248\\
\hline
\end{tabular}
\label{tbl:overhead}
\end{table}

\begin{figure*}
\begin{center}
	\subfloat[Cost-aware Strategy]
	{\label{fig:FDcostDefault}
	\includegraphics[width=0.33\textwidth]
	{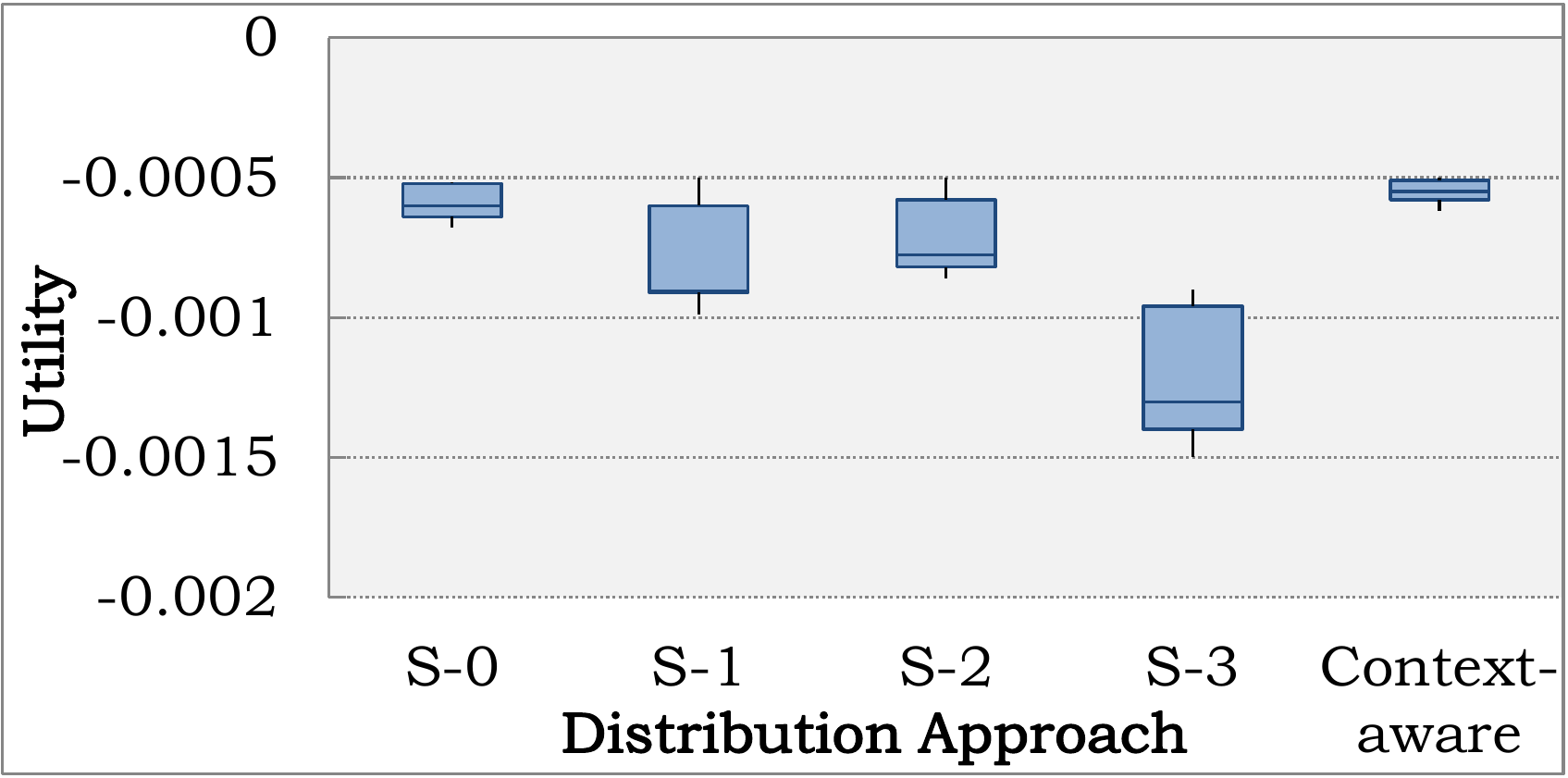}}
	\subfloat[QoS-aware Strategy]
	{\label{fig:FDqosDefault}
	\includegraphics[width=0.33\textwidth]
	{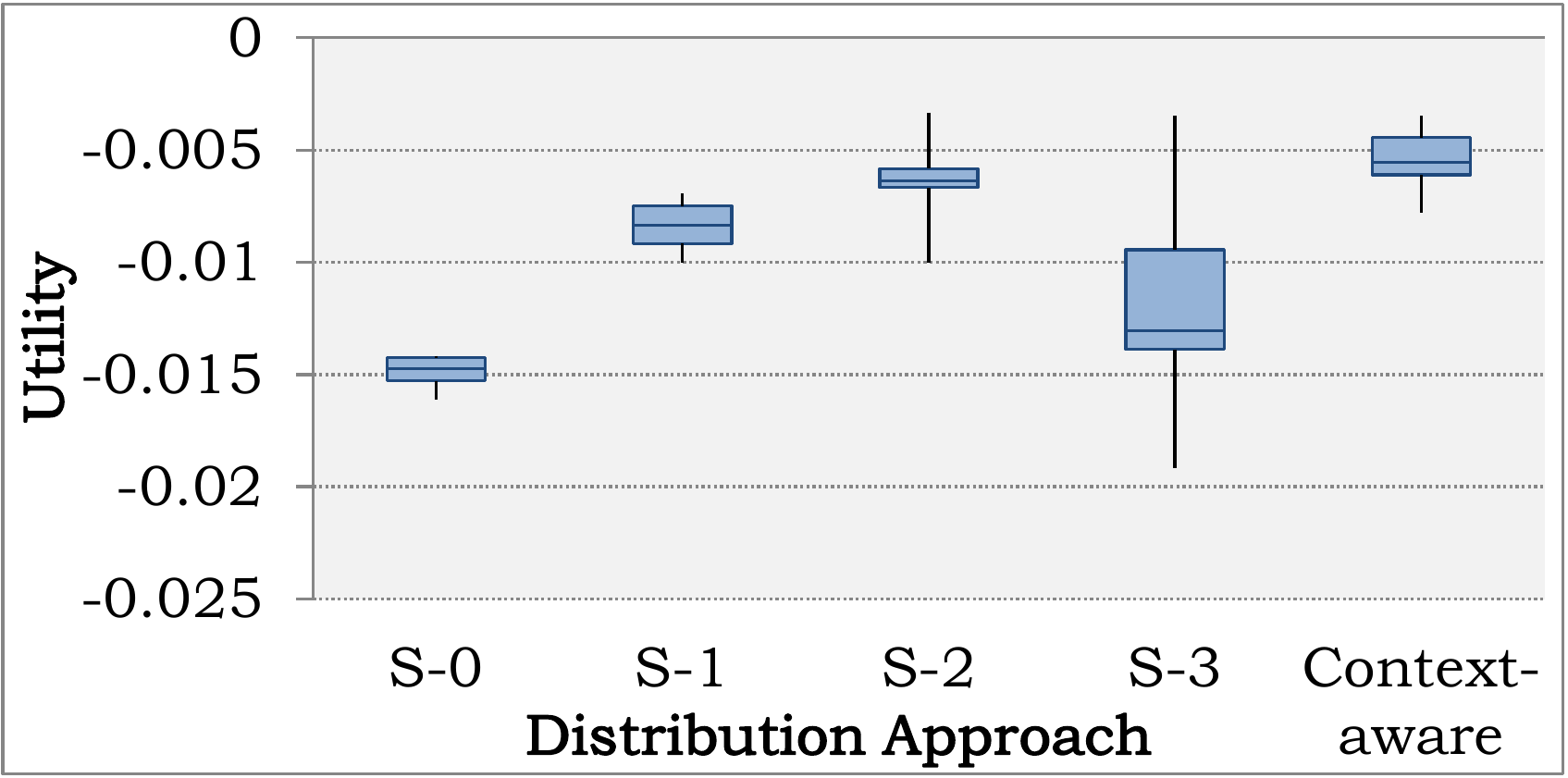}}
	\subfloat[Hybrid Strategy]
	{\label{fig:FDhybridDefault}
	\includegraphics[width=0.33\textwidth]
	{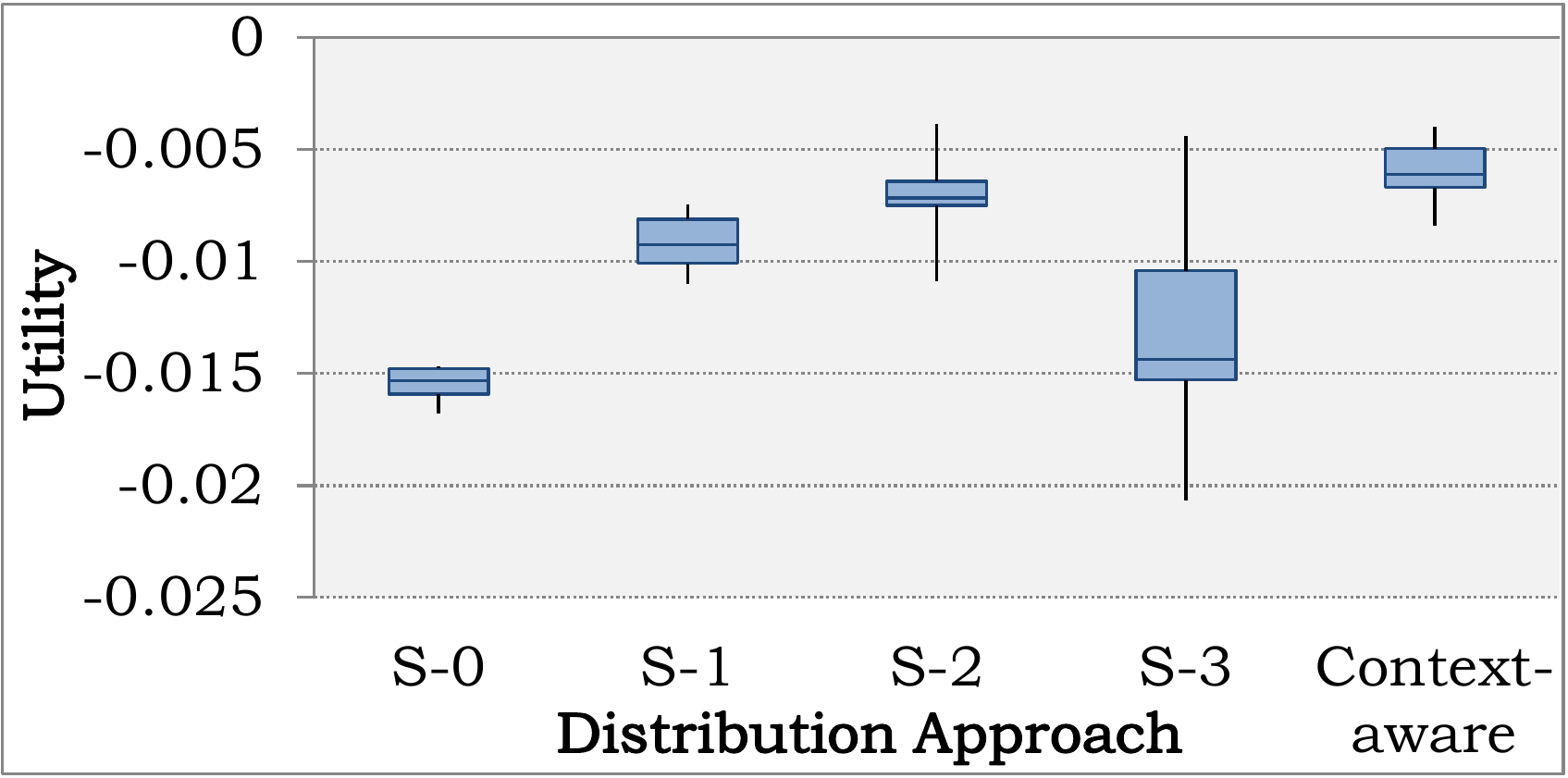}}
\end{center}
\caption{Distribution of utilities over 100 experiments when applying different strategies to the FD application.}
\label{fig:FDperformance}
\end{figure*}

\begin{figure*}
\begin{center}
	\subfloat[Cost-aware Strategy]
	{\label{fig:PMcostDefault}
	\includegraphics[width=0.33\textwidth]
	{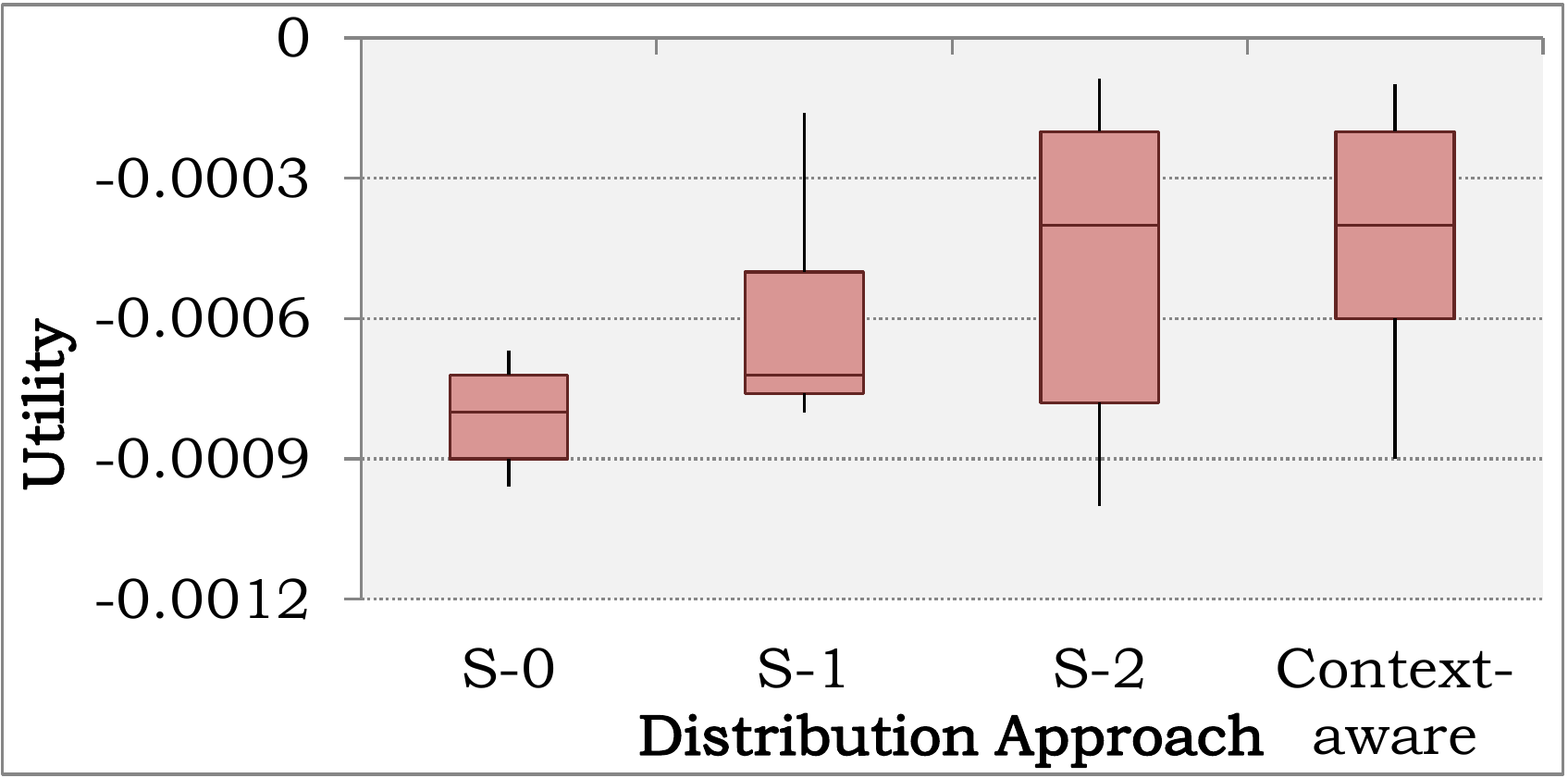}}
	\subfloat[QoS-aware Strategy]
	{\label{fig:PMqosDefault}
	\includegraphics[width=0.33\textwidth]
	{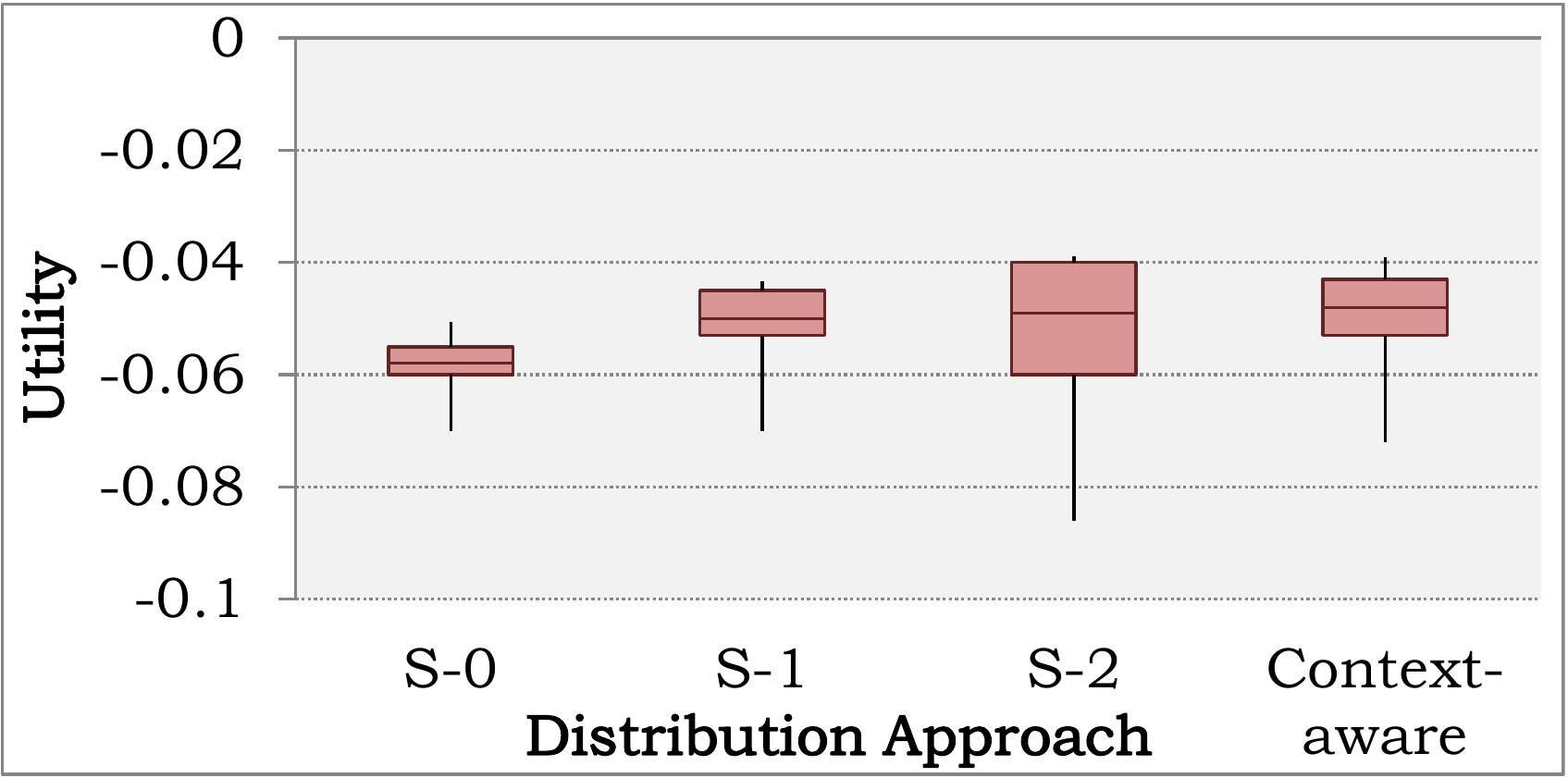}}
	\subfloat[Hybrid Strategy]
	{\label{fig:PMhybridDefault}
	\includegraphics[width=0.33\textwidth]
	{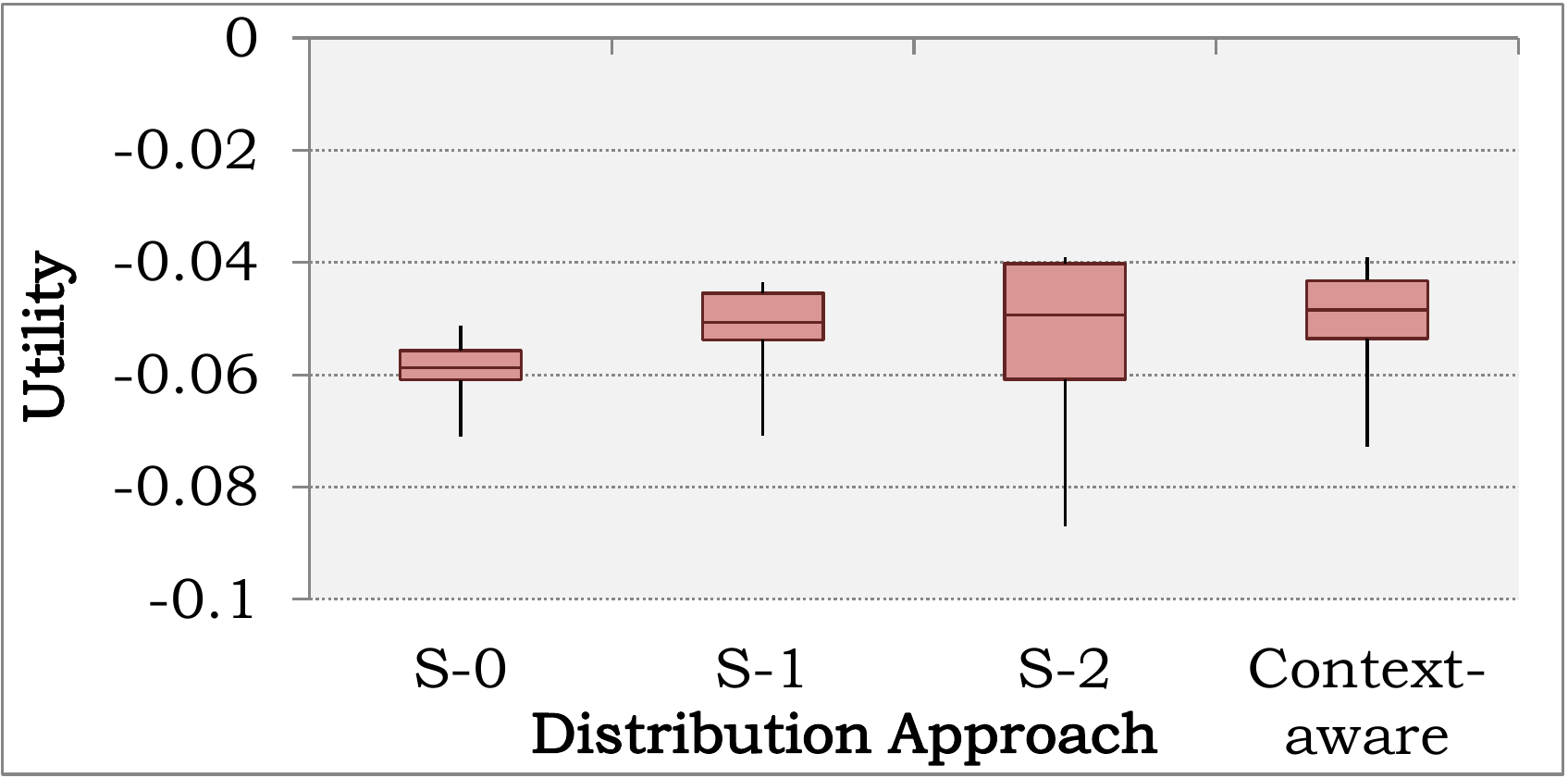}}
\end{center}
\caption{Distribution of utilities over 100 experiments when applying different strategies to the FD application.}
\label{fig:PMperformance}
\end{figure*}

As a multi-tenant computing environment, the resource availability of a Fog node is expected to change rapidly. For example, the system state acquired by the Distribution Manager may become invalid if the context-aware distribution mechanism takes a long time to choose a distribution plan ($k$ value). Therefore, the shorter the time taken by each execution of the decision-making process (Lines~7-26 in Algorithm~\ref{algo:dqn}), the more real-time response is achieved. Each time the Distribution Manager generates a redistribution plan, it takes 85--250 milliseconds (Table~\ref{tbl:overhead}). This overhead may be ignored when compared to the time taken for processing a single video frame using the Cloud-only method, which is nearly 2 seconds from empirical results. In the iPokeMon use-case, this overhead translates into the time taken to process 1--2 iPokeMon requests using the Cloud-only distribution method. Since each deployment of the iPokeMon server is expected to process a large number of user requests, the overhead is negligible.

\subsection{Performance}
\label{sec:performance}

To evaluate the performance of both the static and context-aware approaches, we apply all distribution strategies to 100 new experiments of FD and iPokeMon and display the distribution of their utility values. For context-aware approaches, we took the DQNs trained with 600 and 400 episodes for FD and iPokeMon, respectively, to represent their readily gained knowledge.

Figure~\ref{fig:FDperformance} and~\ref{fig:PMperformance} illustrate the distribution of the utility of each experiment (including 20 deployments) when the static and context-aware approaches are employed for both use-cases. Three combinations of $(\alpha, \beta)$ are considered: (i) $(0, -1)$, which represent a cost-aware strategy (Figure~\ref{fig:FDcostDefault} and~\ref{fig:PMcostDefault}); (ii) $(-1, 0)$, which represents a QoS-aware strategy (Figure~\ref{fig:FDqosDefault} and~\ref{fig:PMqosDefault}); and (iii) the default setting, i.e. $(-1, -1)$, which represent a hybrid strategy (Figure~\ref{fig:FDhybridDefault} and~\ref{fig:PMhybridDefault}).

It is observed for both use-cases that, for all strategies, context-aware approach outperforms the static approaches. For example, when the DQN is trained to minimise cost for FD (Figure~\ref{fig:FDcostDefault}), it achieves the same maximum utility as \textit{S-1} and \textit{S-2}. An improvement over \textit{S-1, S-2} and \textit{S-3} are indicated by the fact that its third quartile of the utility (-0.00058) is larger than the median of the utility obtained by the three static approaches (-0.0006, -0.0009 and -0.0008 respectively). The minimum utility achieves by the DQN is higher than the maximum utility achieves by \textit{S-3}, which means that the cost-optimised DQN never deploy the entire application on the Fog node. This is because, with more modules deployed on the Fog node, there is more cost of using the Fog resources, even though the overall job completion time is reduced.

Similarly, the context-aware approach for iPokeMon (Figure~\ref{fig:PMcostDefault}) tends to benefit from the lessons learnt through \textit{S-1} and \textit{S-2}. It achieves the maximum (-0.00009) and median (-0.0004) utilities that are close to \textit{S-2}, while successfully improves the third quartile (-0.0006) when compared to \textit{S-2} (-0.00078). This is due to the Distribution Manager's selection of \textit{S-1} in a number of deployments.

\begin{figure}
\begin{center}
	\includegraphics[width=0.5\textwidth]
	{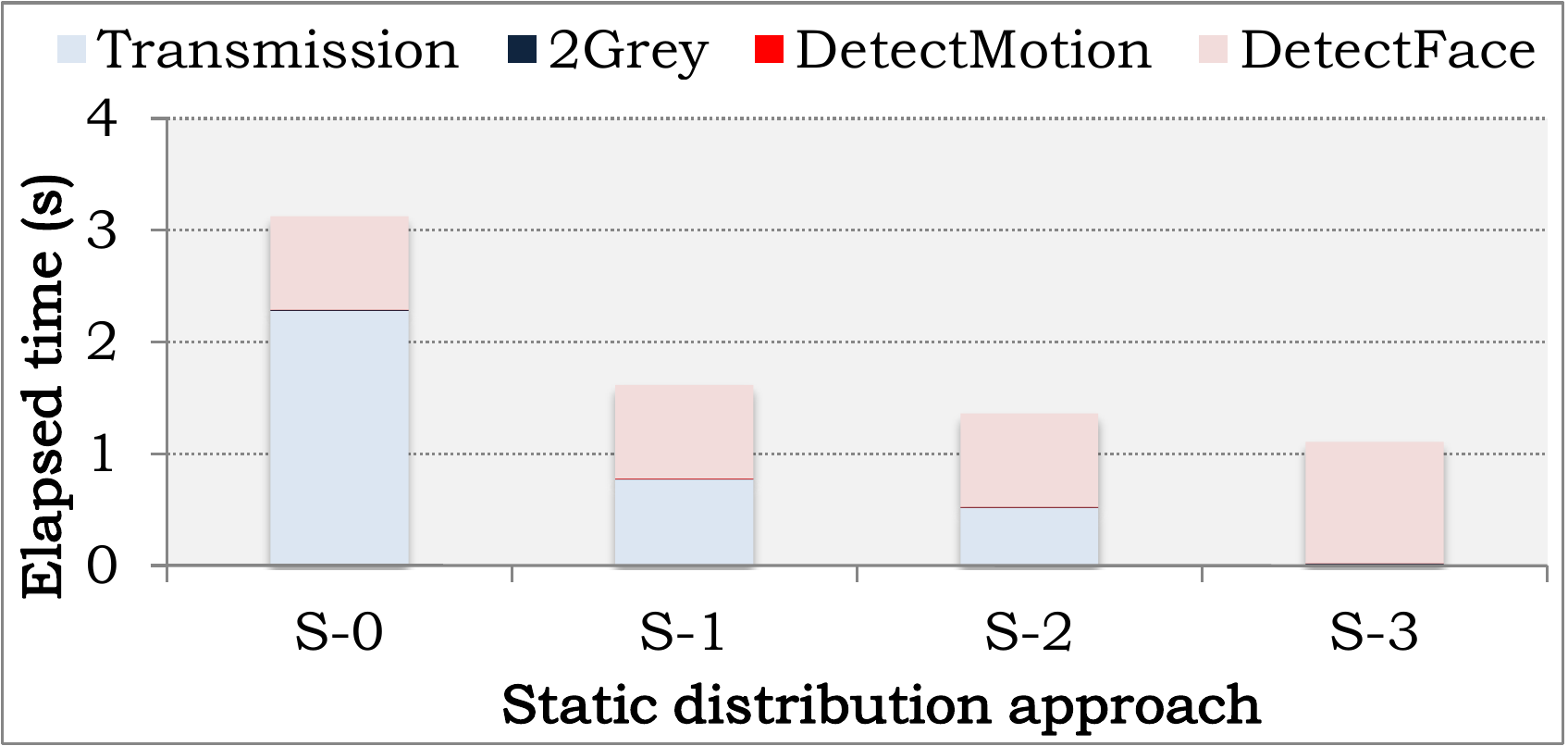}
\end{center}
\caption{Average time taken to process a single video frame in FD using static distribution approaches.}
\label{fig:timeBreakdown}
\end{figure}

The benefit of applying the context-aware approach is more obvious for FD compared to iPokeMon when the DQN is trained to maximise the QoS (i.e. to minimise the job completion time, Figure~\ref{fig:FDqosDefault}). The majority of the utilities achieved by the context-aware approach is larger than that achieved by all four static approaches. To better understand the difference of the QoS when applying different static approaches, Figure~\ref{fig:timeBreakdown} provides a breakdown of the average time to process a single video frame when the Fog node system is not stress-tested (i.e. almost all CPU cores and memory are available). The overall time is divided into the data transmission time, the time taken by the three modules of the application -- grey-scale conversion, motion detection and face detection. It is inferred that the main difference of the QoS comes from the data transmission time. For example, by applying $S-1, S-2$ and $S-3$, the transmission time is reduced from 2.28 seconds (s) to 0.77s, 0.52s and 0.11s respectively. The first two modules of the application, namely grey-scale conversion and motion detection take a short time between 0.003 and 0.004, no matter where they are hosted. The face detection module causes the other main difference among these approaches as there is 0.2s delay observed when it is hosted on the Fog node instead of the Cloud VM. Such differences of QoS among the static approaches are expected to be magnified in the experiment for Figure~\ref{fig:FDqosDefault} when the Fog resources are deliberately restricted. Therefore, by in the context-aware approach, the DQN tends to only choose the optimal deployment from \textit{S-1}, \textit{S-2} and \textit{S-3} to avoid the long transmission time caused by \textit{S-0}.  

When applying the QoS-aware strategy to iPokeMon~\ref{fig:PMqosDefault}, the context-aware approach improves the overall application performance over the static approaches by achieving the highest median (-0.048) and third quartile (-0.053) values of utility. However, in the best and worst cases, the context-aware approach performs slightly worse than \textit{S-3} by 0.3\% and \textit{S-1} by 2\%. This difference is indicative that the context-aware approach has more benefits for FD~\ref{fig:FDqosDefault} since there is a larger QoS gain over iPokeMon.

When we consider both the cost and the QoS (Figure~\ref{fig:FDhybridDefault} and~\ref{fig:PMhybridDefault}), the distribution of all approaches for both use-cases is similar to the QoS-aware strategy. The reason is that with the default parameter setting, the difference of QoS among the approaches happens to have a larger influence than the difference of costs. Therefore, in the next section, we further investigate the impact of different parameter settings on the utility.

\subsection{Impact of Utility Parameters
}
\label{sec:parameter}
From empirical study, we found that when assigning different weights to $\alpha$ and $\beta$ (i.e. the relative importance of QoS and cost), the performance of all approaches varies. Another factor that makes a difference is $\lambda$, which indicates how expensive the Fog resources are priced compared to the Cloud resources. Hence, several different values for $(\alpha, \beta)$ and $\lambda$ are applied to explore their impacts on the distribution approaches.

\begin{figure}
\begin{center}
	\subfloat[FD]
	{\label{fig:FDlambda}
	\includegraphics[width=0.5\textwidth]
	{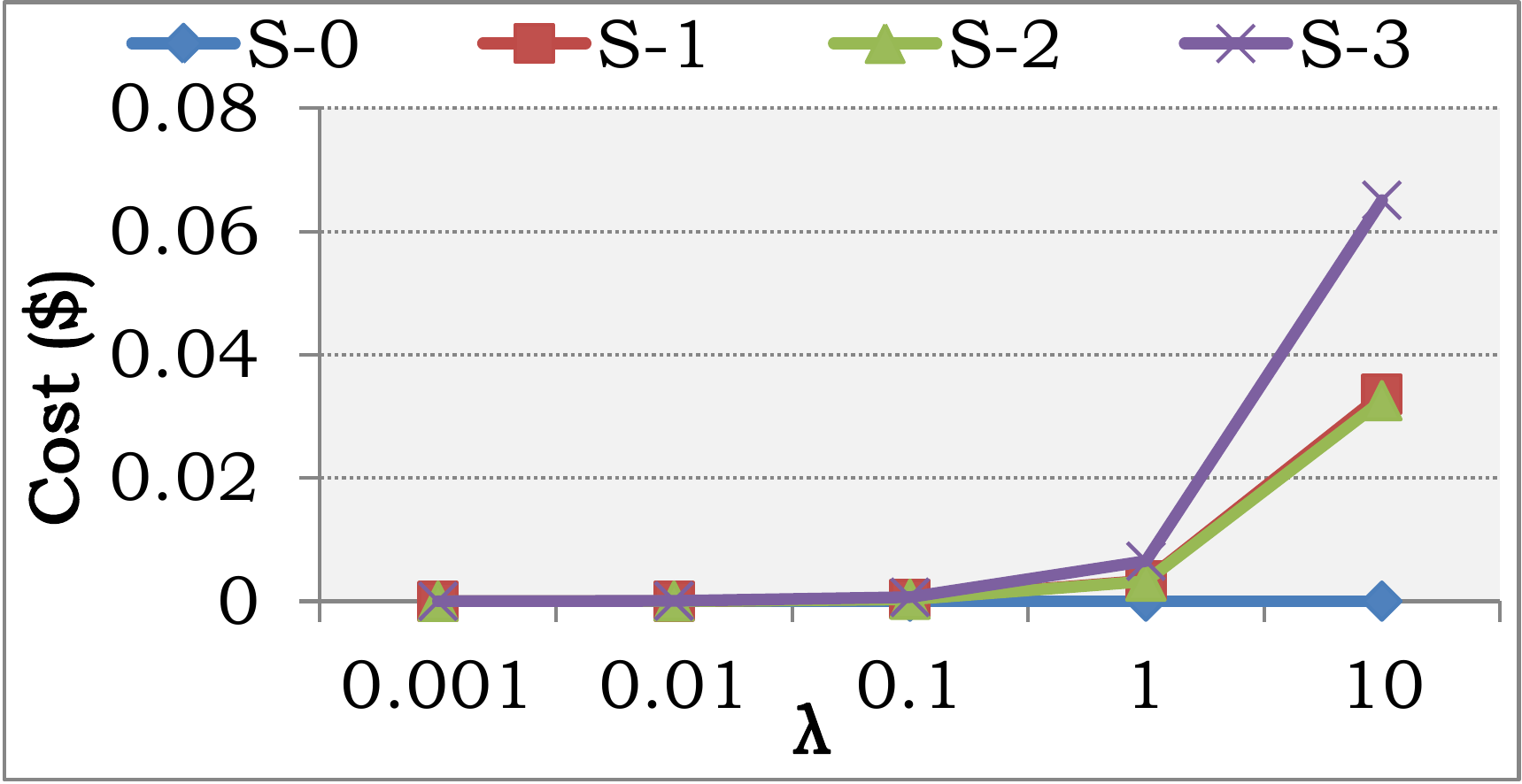}}
	\hfill
	\subfloat[iPokeMon]
	{\label{fig:PMlambda}
	\includegraphics[width=0.5\textwidth]
	{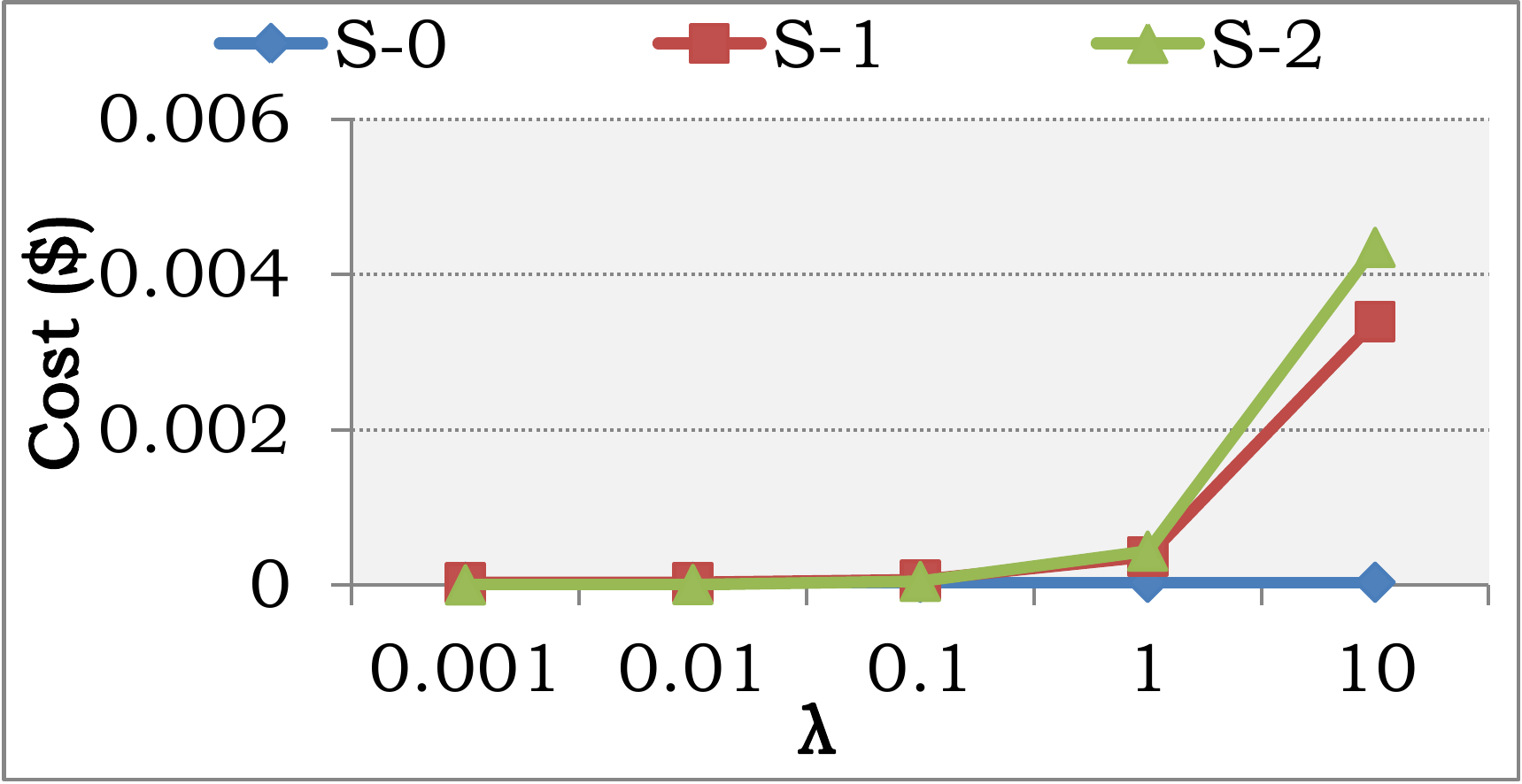}}
\end{center}
\caption{Average cost of a single deployment using static distribution approaches with varying $\lambda$.}
\label{fig:lambda}
\end{figure}

Figure~\ref{fig:lambda} presents the relationship between the average cost of a single deployment and $\lambda$ for FD and iPokeMon when the Fog node system is not stress tested. It is noted that in the FD use case although the Fog computing services effectively improves the QoS (Figure~\ref{fig:timeBreakdown}), its cost increases drastically when $\lambda$ is larger than 0.1~\ref{fig:FDlambda}. For example, the cost of \textit{S-3} is as 200 and 2,000 times large as the cost of \textit{S-0}, when $\lambda$ is 1 and 10 respectively. In such cases, when $\beta \neq 0$ the context-aware approach would always tend to choose \textit{S-1} in order to minimise the cost factor in Equation~\ref{eq:utility}. This would not distinguish the context-aware approach from the static approaches. For the iPokeMon use case~\ref{fig:PMlambda}, the increase in cost is moderate. For instance, the cost of \textit{S-2} is close to 0.1, 1.3, and 13 times of \textit{S-0} when $\lambda$ is 0.01, 0.1 and 1 respectively. Therefore, in the following experiments we consider $\lambda \in \{0.0001, 0.001, 0.1\}$ for FD and $\lambda \in \{0.001, 0.1, 1\}$ for iPokeMon to explore the impact of $\lambda$ over the application performances. We also present the analysis of varying importance assigned to the QoS and cost factors over the application performances, through applying $\lambda/\beta \in \{0.1, 1, 10\}$ for both use-cases.

\begin{figure}
\begin{center}
	\subfloat[$\lambda = 0.1$]
	{\label{fig:FDcostLambda0.1}
	\includegraphics[width=0.5\textwidth]
	{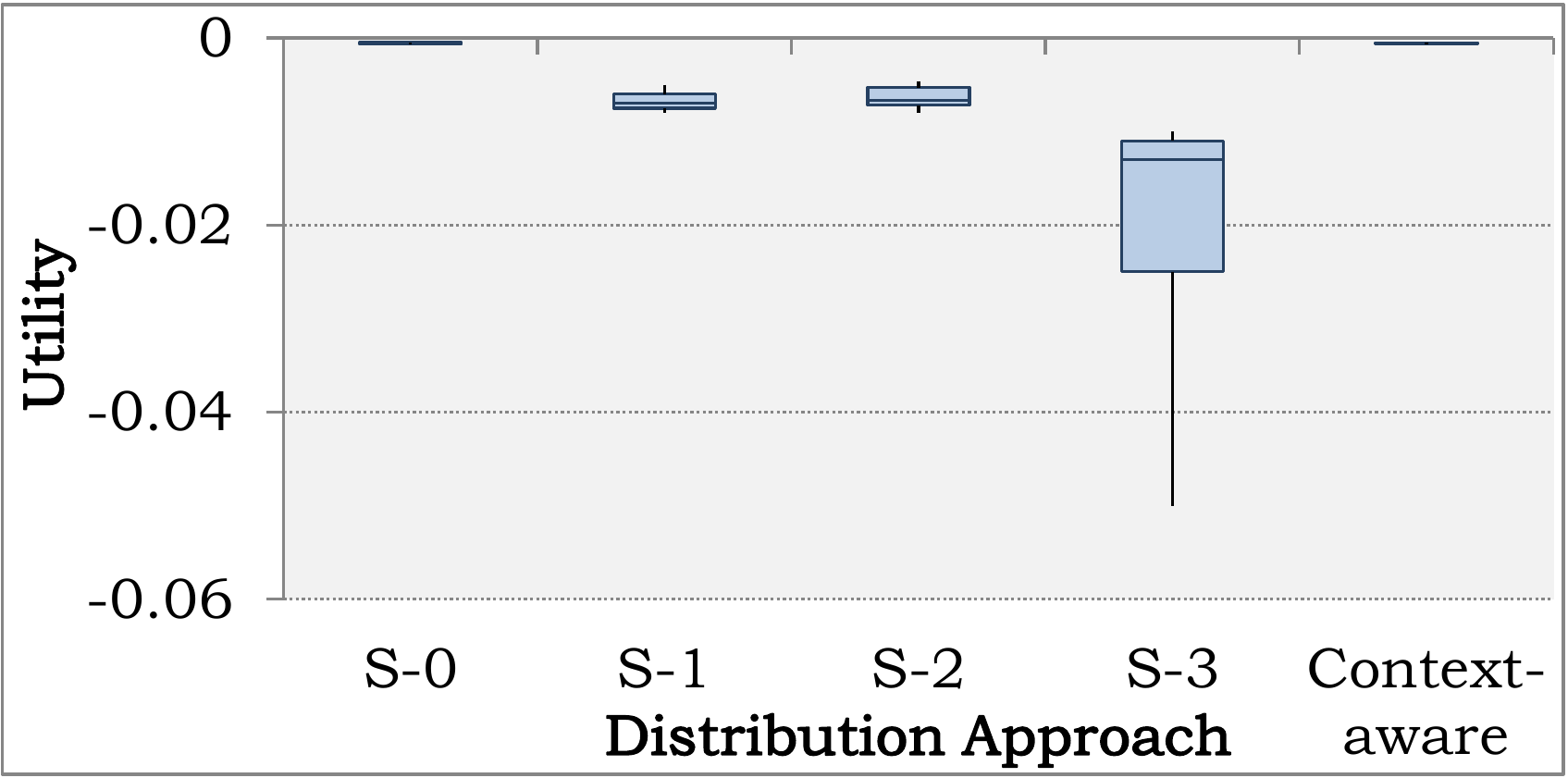}}
	\hfill
	\subfloat[$\lambda = 0.001$]
	{\label{fig:FDcostLambda0.001}
	\includegraphics[width=0.5\textwidth]
	{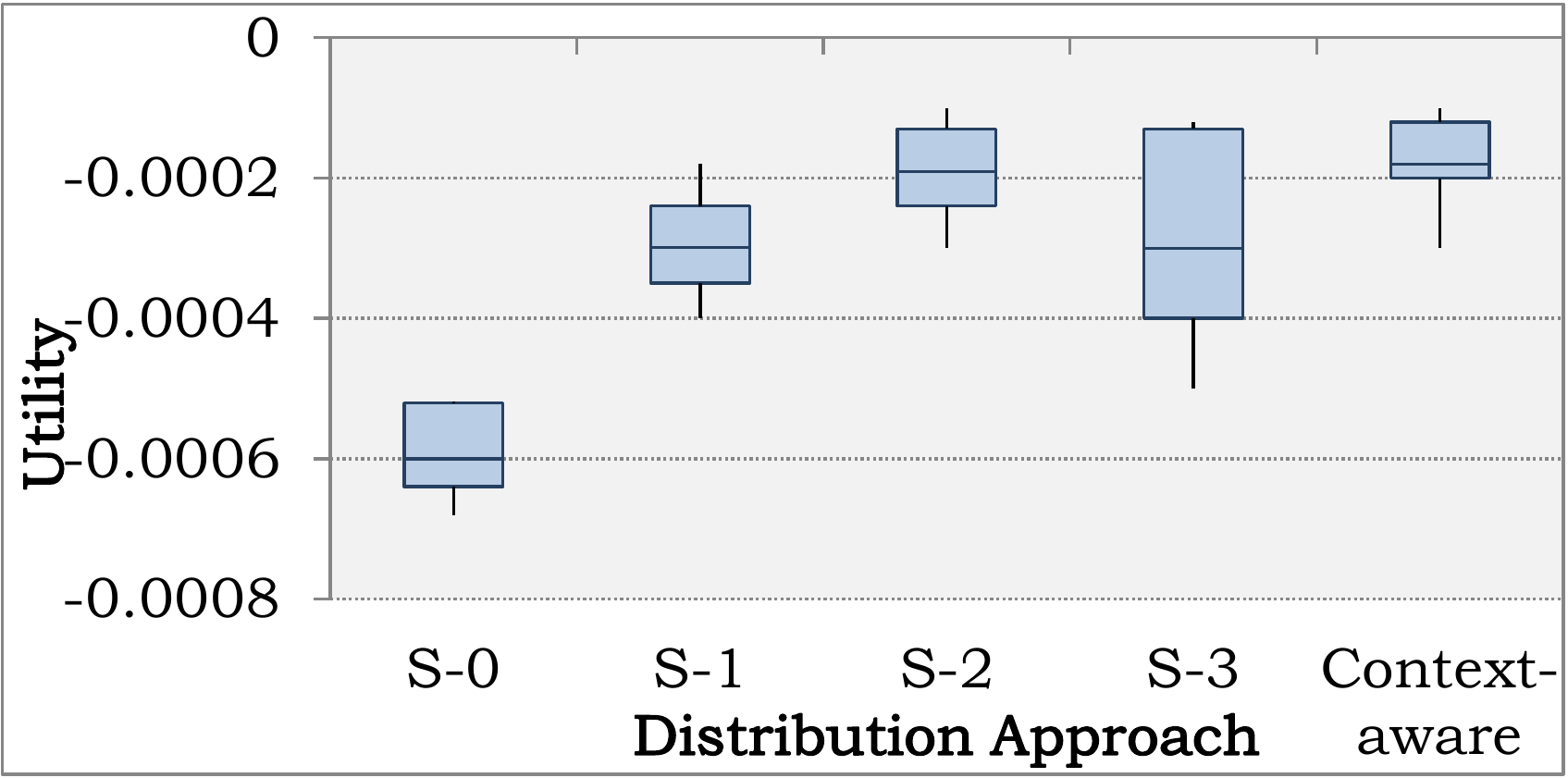}}
\end{center}
\caption{Distribution of utilities over 100 experiments when $\alpha / \beta = 0$ with varying $\lambda$ for the FD application.}
\label{fig:FDab0}
\end{figure}

\begin{figure}
\begin{center}
	\subfloat[$\lambda = 1$]
	{\label{fig:PMcostLambda1}
	\includegraphics[width=0.5\textwidth]
	{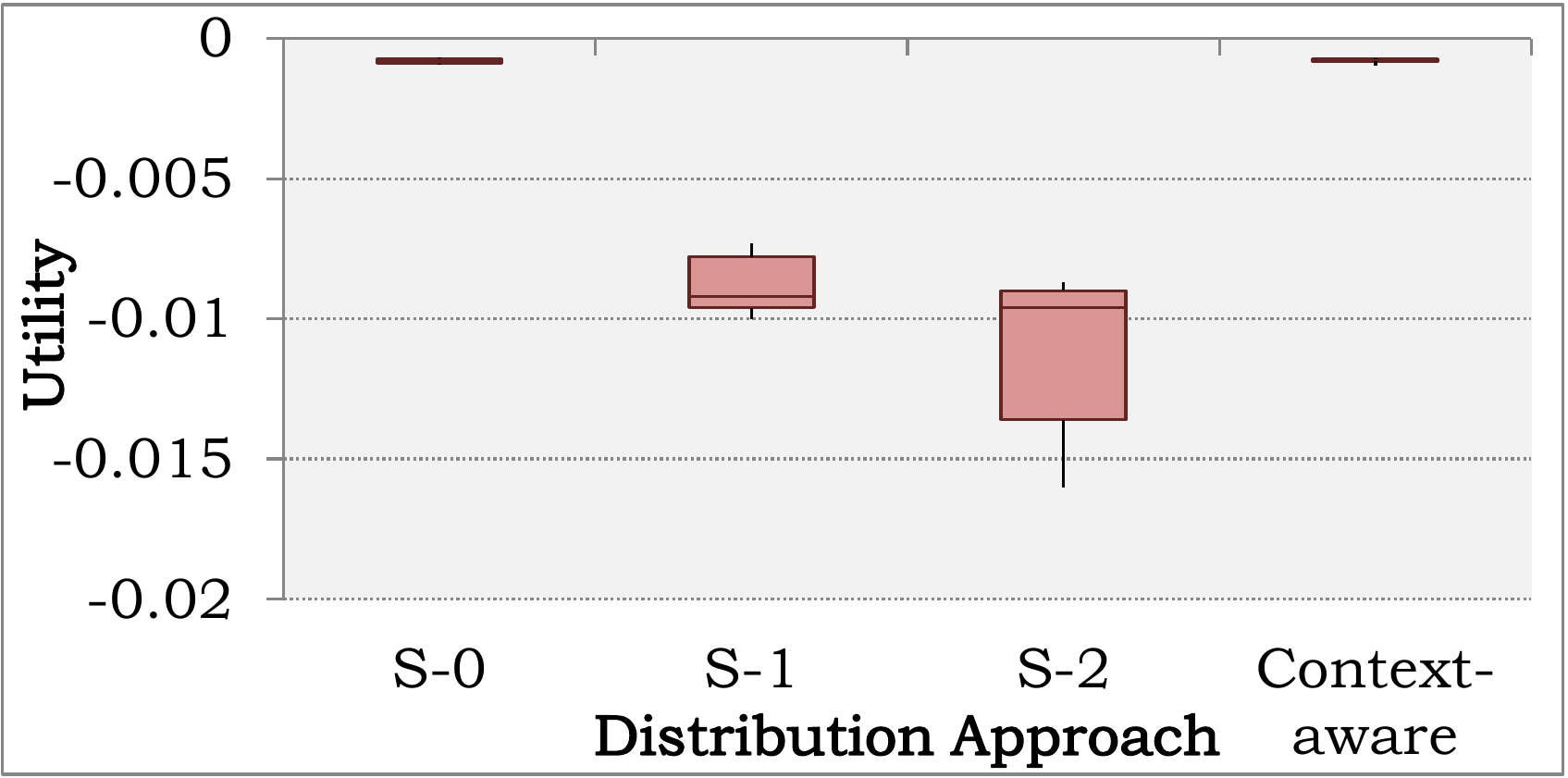}}
	\hfill
	\subfloat[$\lambda = 0.1$]
	{\label{fig:PMcostLambda0.1}
	\includegraphics[width=0.5\textwidth]
	{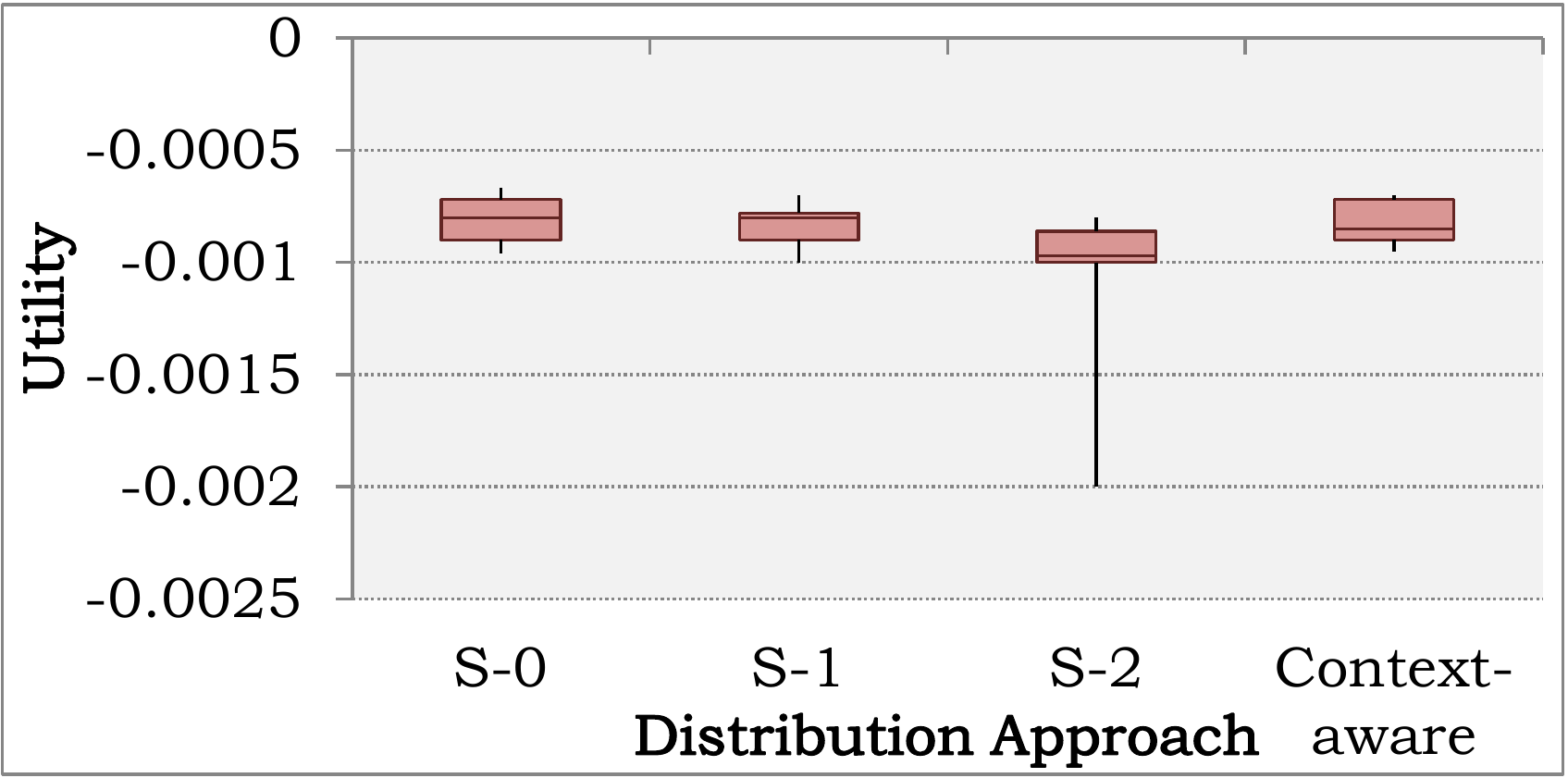}}
\end{center}
\caption{Distribution of utilities over 100 experiments when $\alpha / \beta = 0$ with varying $\lambda$ for iPokeMon.}
\label{fig:PMab0}
\end{figure}

\begin{figure*}
\begin{center}
	\subfloat[$\lambda = 0.1$]
	{\label{fig:FDhDot1HighLambda}
	\includegraphics[width=0.33\textwidth]
	{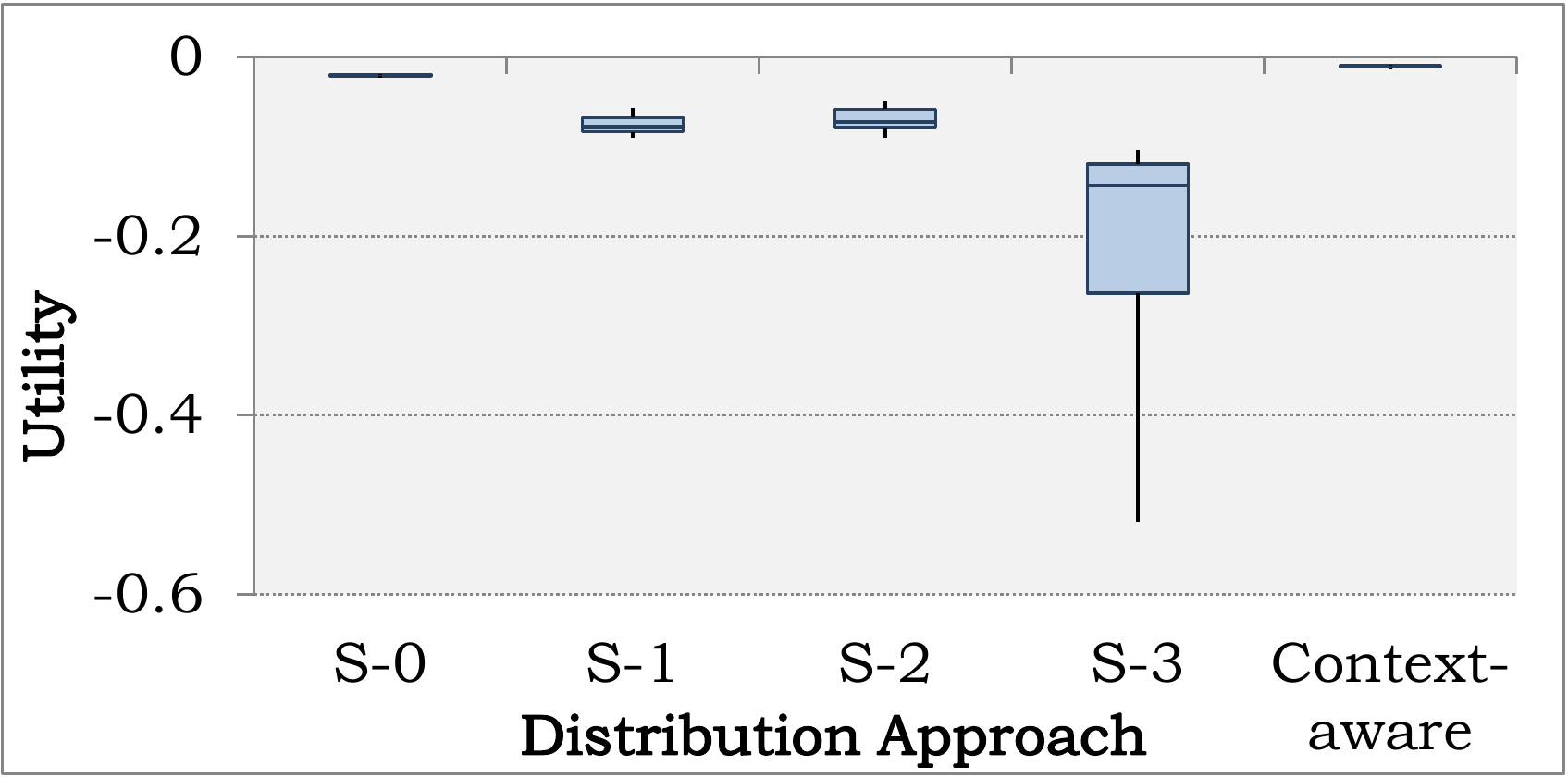}}
	\subfloat[$\lambda = 0.01$]
	{\label{fig:FDhDot1MediumLambda}
	\includegraphics[width=0.33\textwidth]
	{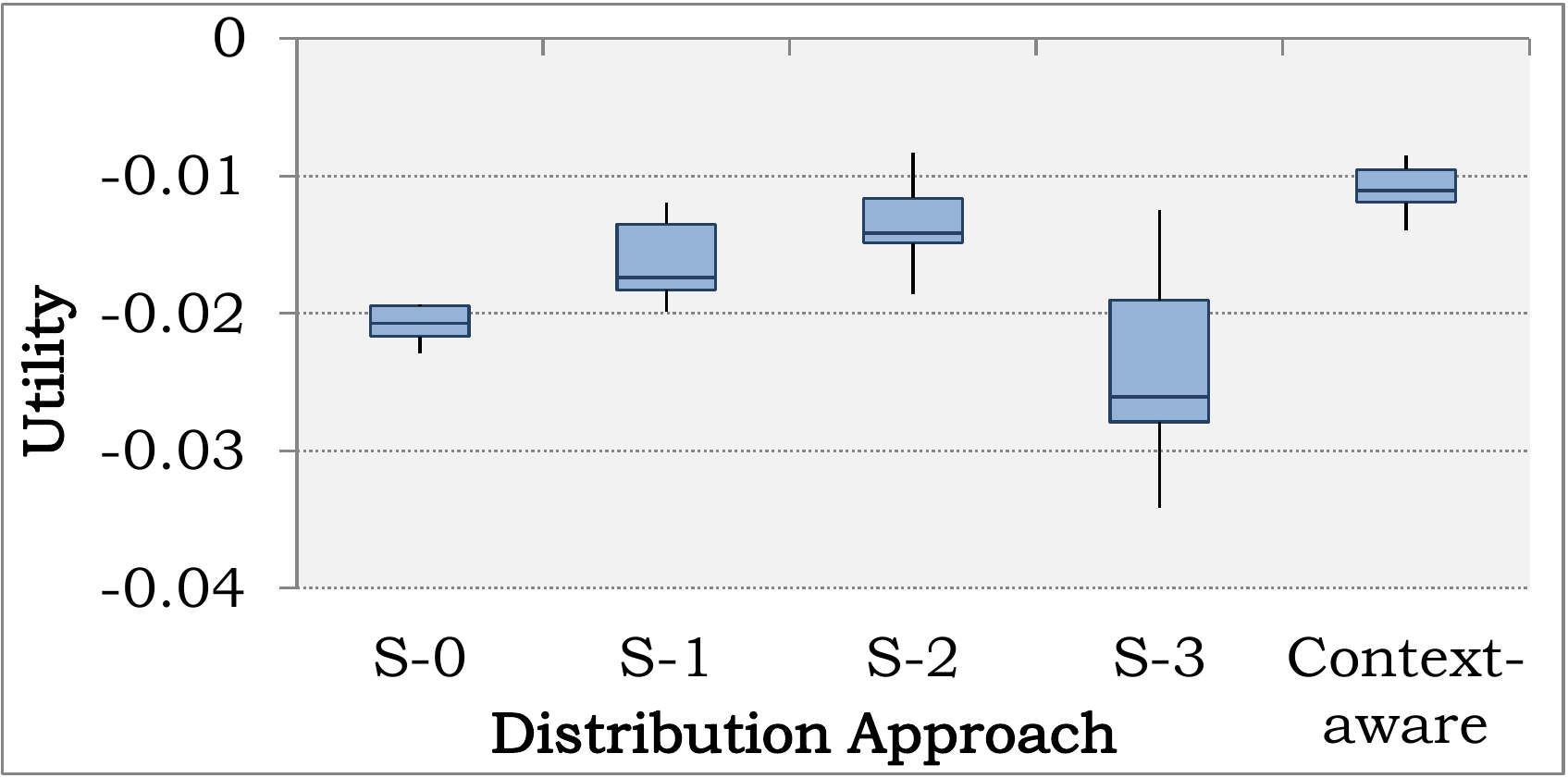}}
	\subfloat[$\lambda = 0.001$]
	{\label{fig:FDhDot1LowLambda}
	\includegraphics[width=0.33\textwidth]
	{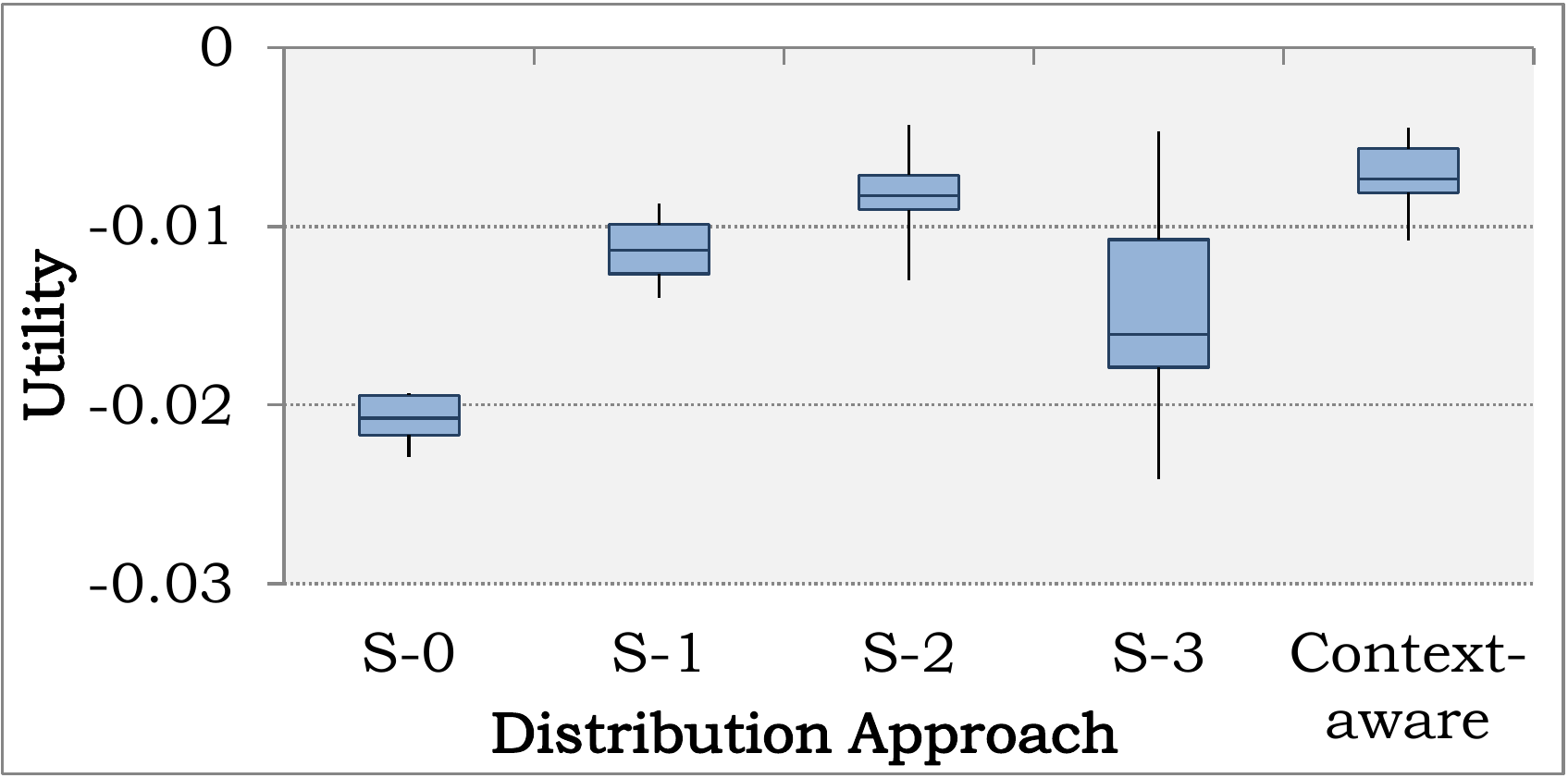}}
\end{center}
\caption{Distribution of utilities over 100 experiments when $\alpha / \beta = 0.1$ with varying $\lambda$ for the FD application.}
\label{fig:FDhybrid0.1}
\end{figure*} 

\begin{figure*}
\begin{center}
	\subfloat[$\lambda = 1$]
	{\label{fig:PMhDot1HighLambda}
	\includegraphics[width=0.33\textwidth]
	{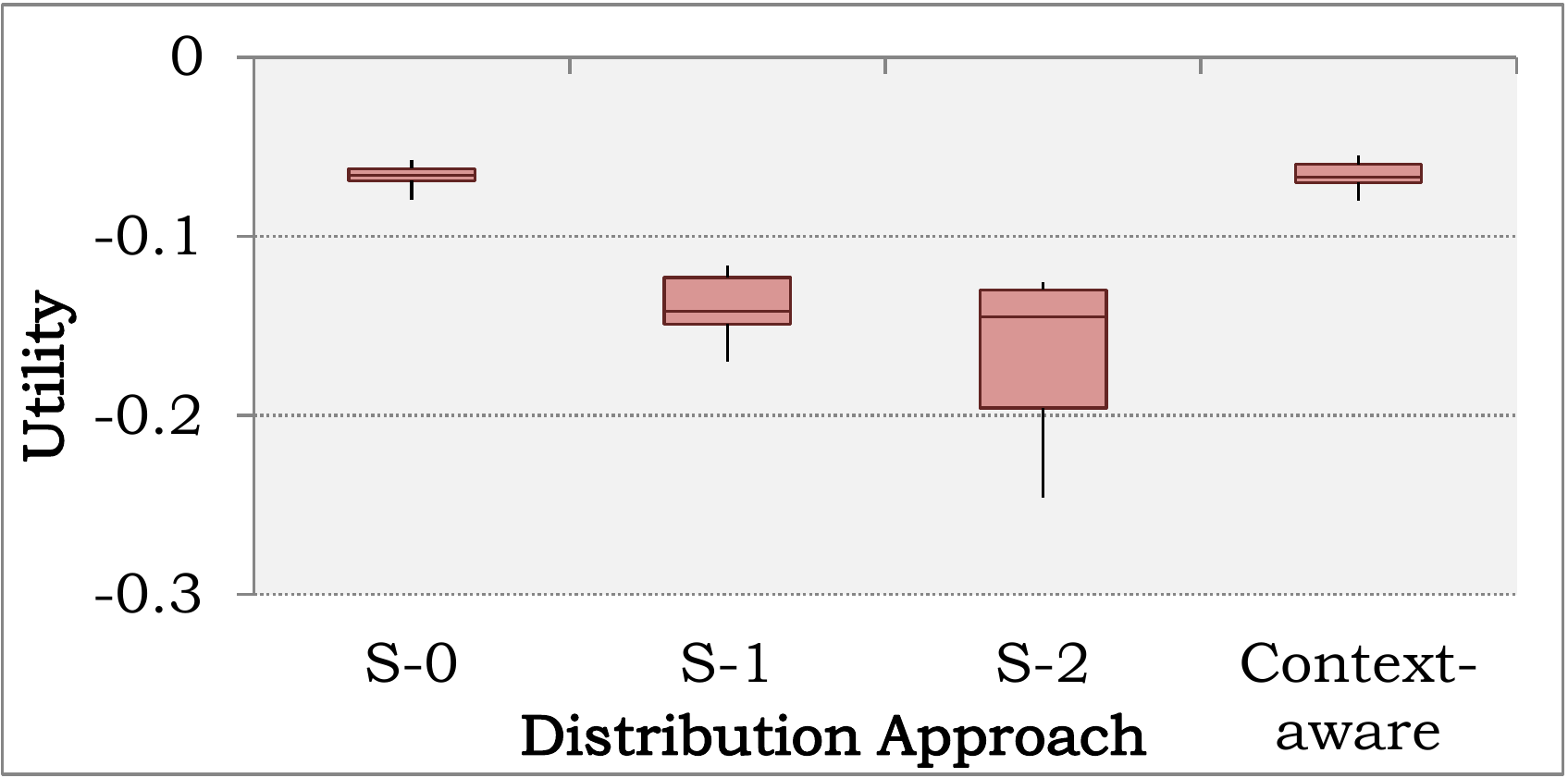}}
	\subfloat[$\lambda = 0.1$]
	{\label{fig:PMhDot1MediumLambda}
	\includegraphics[width=0.33\textwidth]
	{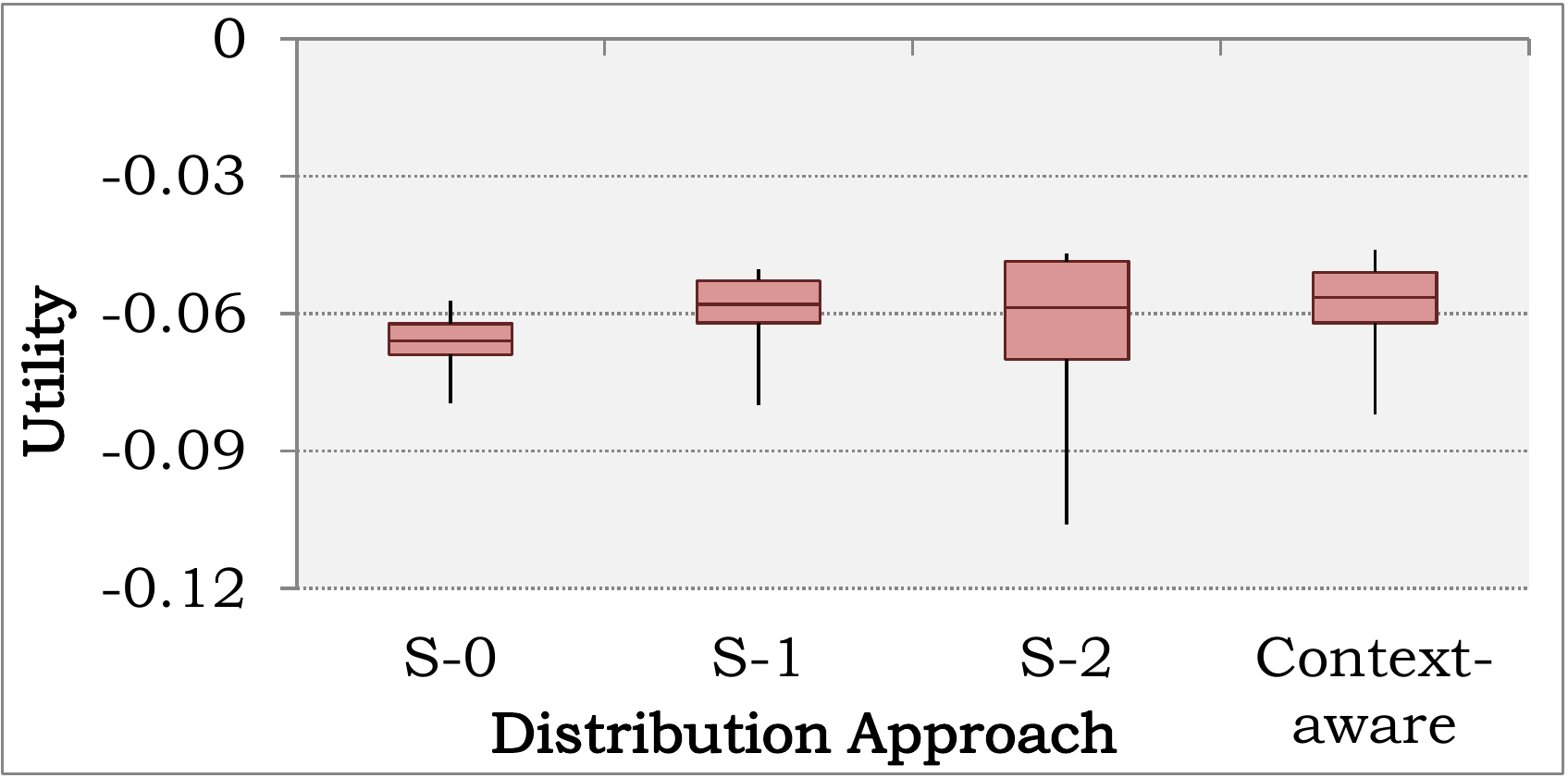}}
	\subfloat[$\lambda = 0.01$]
	{\label{fig:PMhDot1LowLambda}
	\includegraphics[width=0.33\textwidth]
	{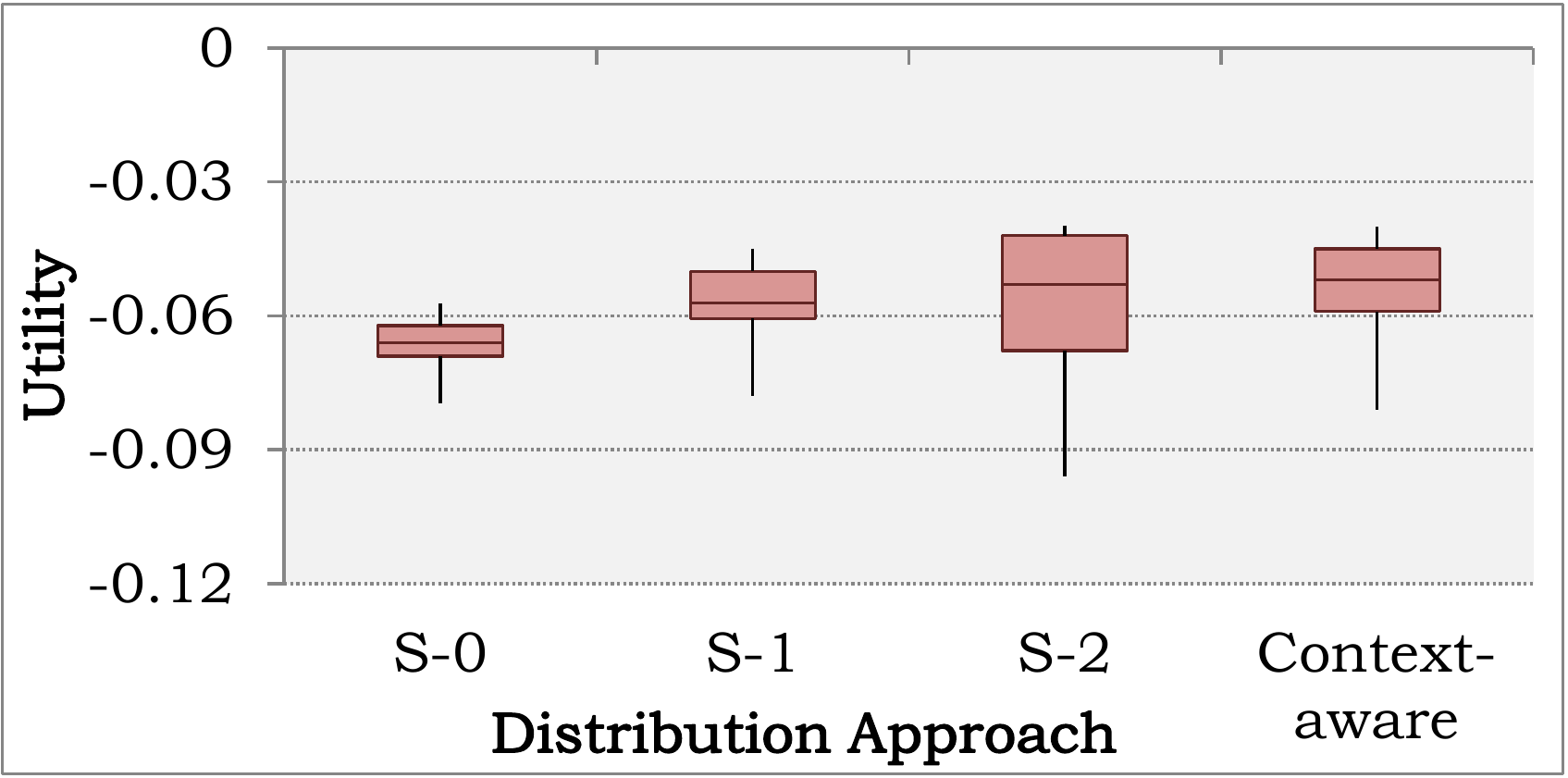}}
\end{center}
\caption{Distribution of utilities over 100 experiments when $\alpha / \beta = 0.1$ with varying $\lambda$ for iPokeMon.}
\label{fig:PMhybrid0.1}
\end{figure*}

Figure~\ref{fig:FDab0} displays the performance of different approaches when $\alpha/\beta =0$ (i.e. the cost-efficient strategy) with the higher and lower values of $\lambda$ in addition to the medium $\lambda$ (Figure~\ref{fig:FDcostDefault}). The Fog resources are randomly restricted as in Section~\ref{sec:performance}. When $\lambda = 0.1$, i.e. Fog resources are priced as one-tenth of Cloud resources, the context-aware approach acts the same as \textit{S-0} since \textit{S-0} has a clear advantage over the other static approaches. When $\lambda = 0.001$, i.e. Fog resources are priced as one-thousandth of Cloud resources, the context-aware approach can benefit from the Fog services. Note that under such an assumption the overall costs of Fog-based distribution approaches (\textit{S-1, S-2} and \textit{S-3}) become less than the Cloud-only distribution approach (\textit{S-0}). When $\lambda = 0.01$ (Figure~\ref{fig:FDcostDefault}), i.e. Fog resources are priced as one hundredth of Cloud resources, the context-aware approach benefits from some of the Fog-based distribution approaches (\textit{S-1} and \textit{S-2}). In this case, the costs of Fog-based distribution approaches are slightly more than the Cloud-only distribution approach. 

Figure~\ref{fig:PMab0} displays the performance of iPokeMon when different approaches are applied with the cost-efficient strategy and with the higher and medium values of $\lambda$ in addition to the lower $\lambda$ (Figure~\ref{fig:PMcostDefault}). When $\lambda = 1$, i.e. Fog resources are priced the same as Cloud resources, the context-aware approach acts the same as \textit{S-0} since \textit{S-1} has a clear advantage over the other static approaches. When $\lambda = 0.1$, the distribution of utility in the context-aware approach is close to \textit{S-0}, with a variance in the median value caused by the occasional selection of \textit{S-2} and \textit{S-3}. When $\lambda = 0.01$ (Figure~\ref{fig:PMcostDefault}), the context-aware approach mainly benefits from the Fog-based, and Fog-only distribution approaches as the costs of these deployment plans are less than the Cloud-only distribution approach.

\begin{figure}
\begin{center}
	\subfloat[$\lambda = 0.1$]
	{\label{fig:FDh1HighLambda}
	\includegraphics[width=0.5\textwidth]
	{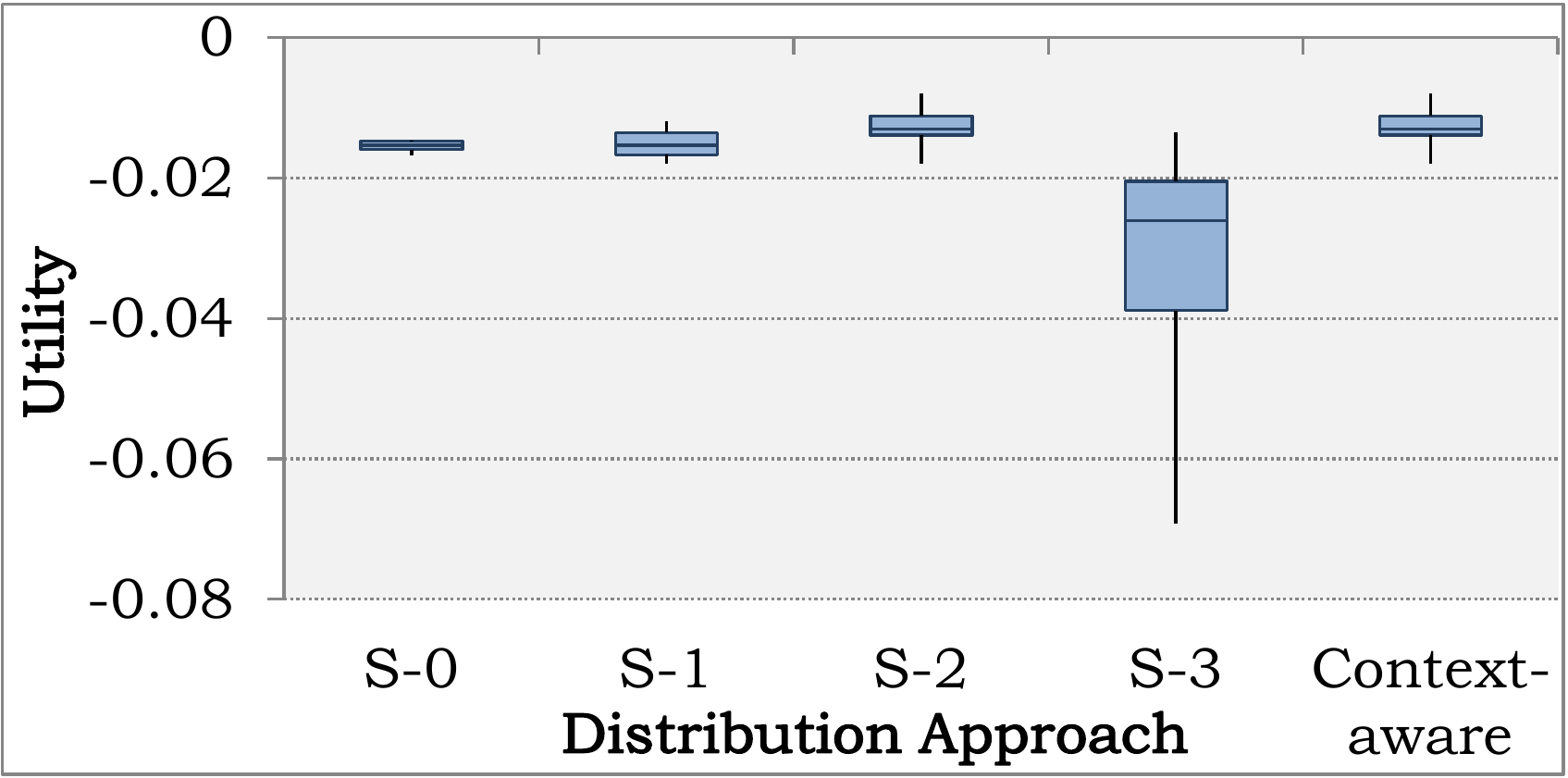}}
	\hfill
	\subfloat[$\lambda = 0.001$]
	{\label{fig:FDh1LowLambda}
	\includegraphics[width=0.5\textwidth]
	{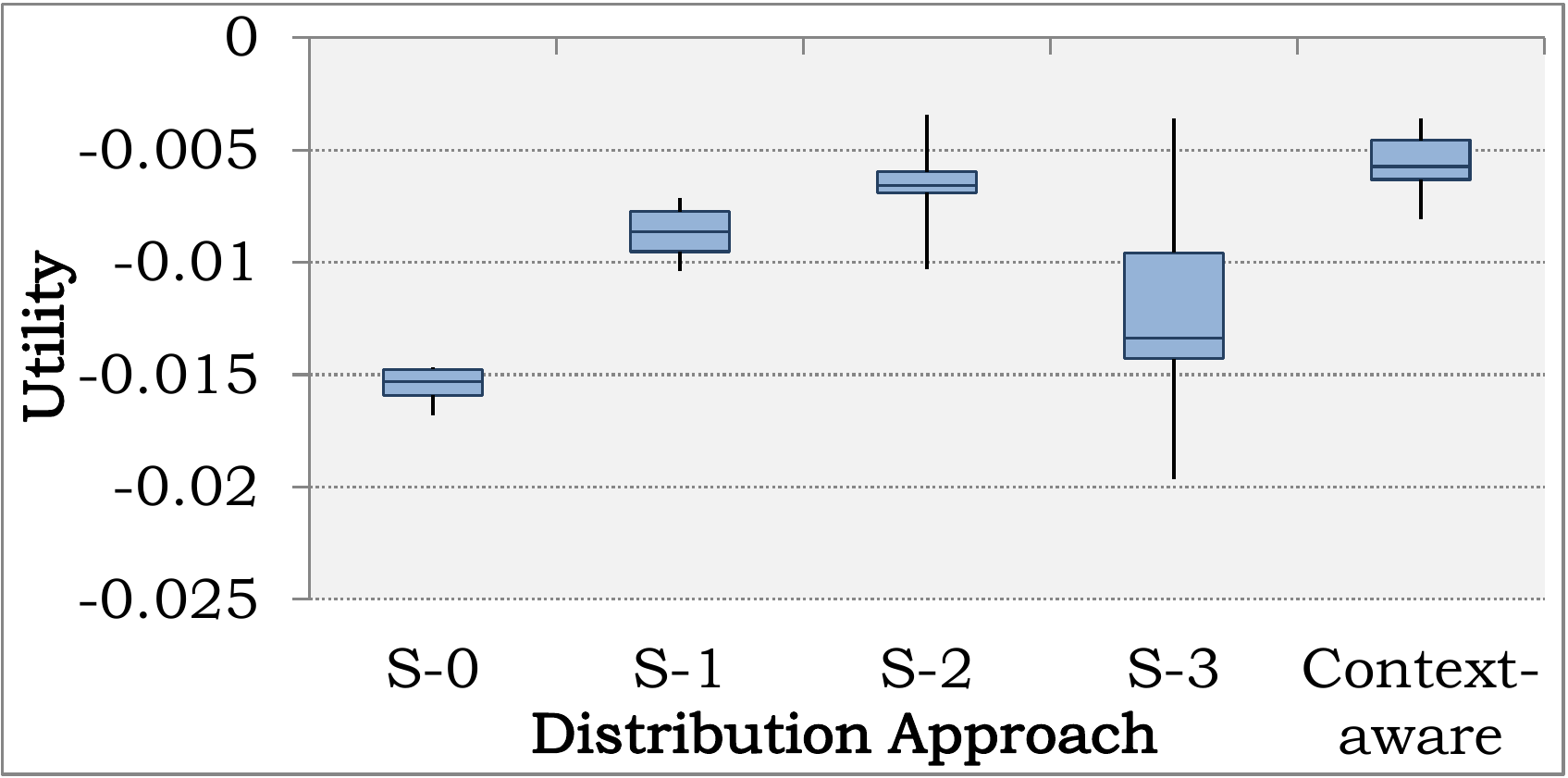}}
\end{center}
\caption{Distribution of utilities over 100 experiments when $\alpha / \beta = 1$ with varying $\lambda$ for the FD application.}
\label{fig:FDhybrid1}
\end{figure}

\begin{figure}
\begin{center}
	\subfloat[$\lambda = 1$]
	{\label{fig:PMh1HighLambda}
	\includegraphics[width=0.5\textwidth]
	{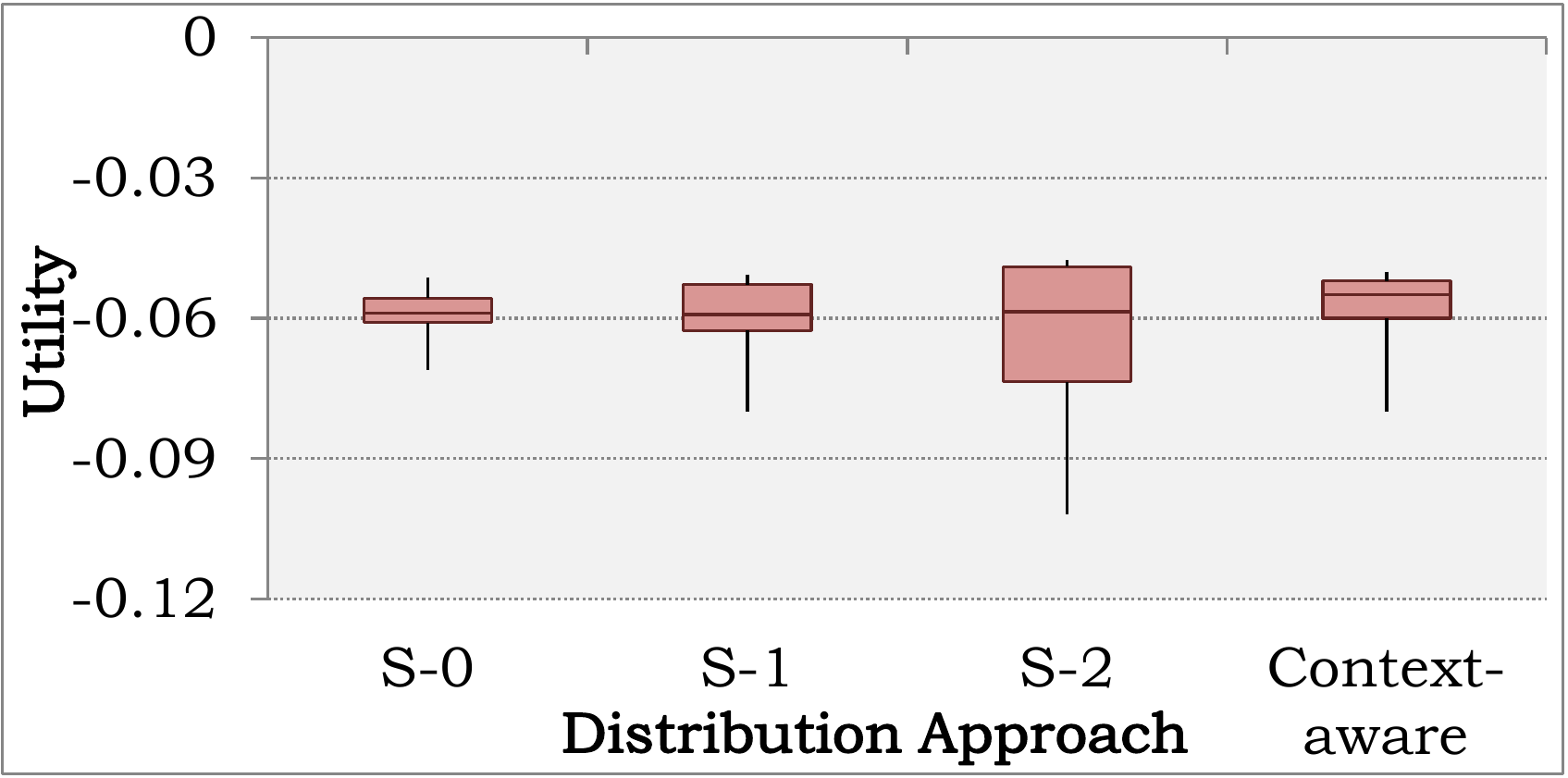}}
	\hfill
	\subfloat[$\lambda = 0.1$]
	{\label{fig:PMh1MediumLambda}
	\includegraphics[width=0.5\textwidth]
	{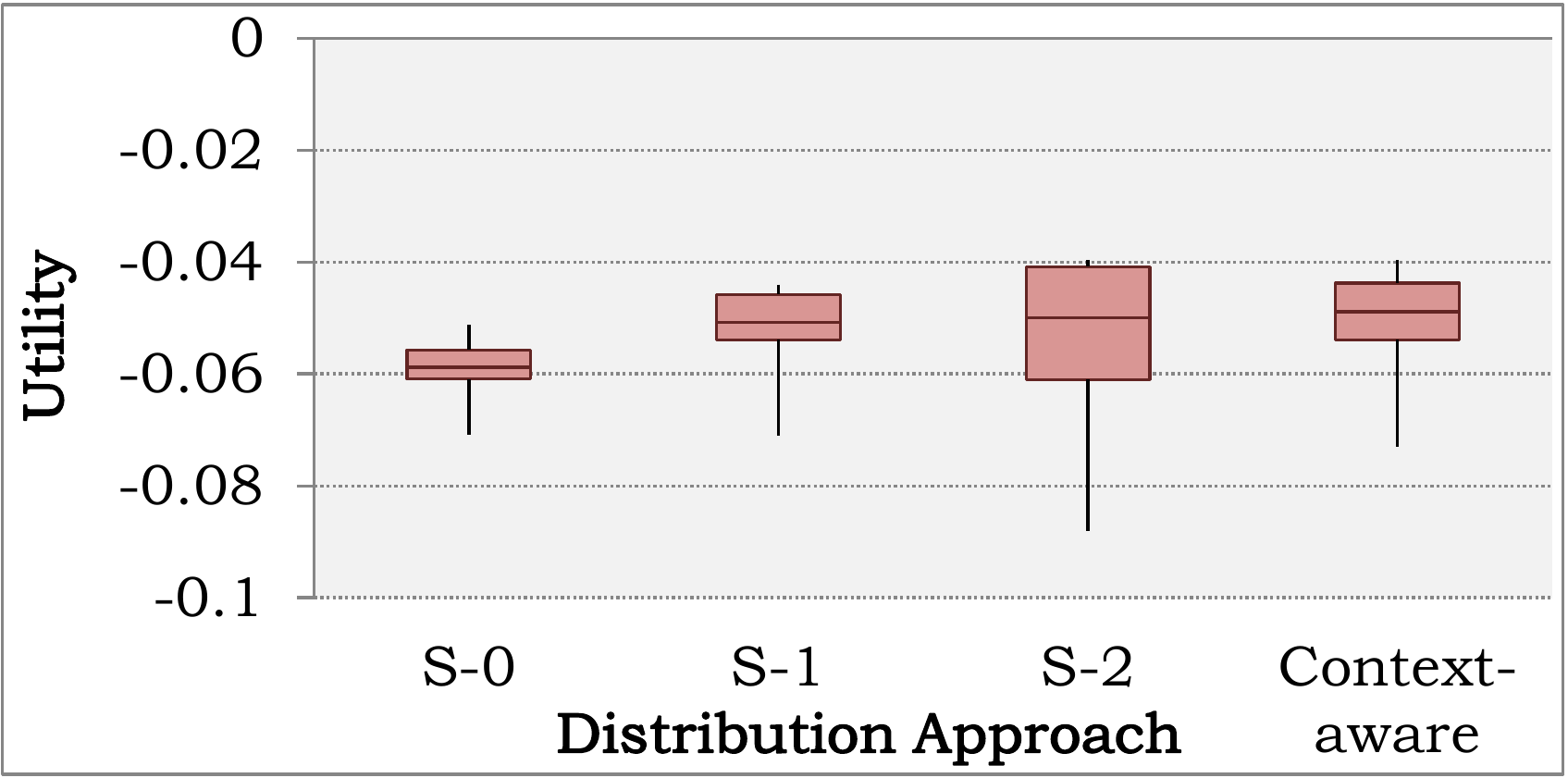}}
\end{center}
\caption{Distribution of utilities over 100 experiments when $\alpha / \beta = 1$ with varying $\lambda$ for iPokeMon.}
\label{fig:PMhybrid1}
\end{figure} 

\begin{figure*}
\begin{center}
	\subfloat[$\lambda = 0.1$]
	{\label{fig:FDh10Lambda0.1}
	\includegraphics[width=0.33\textwidth]
	{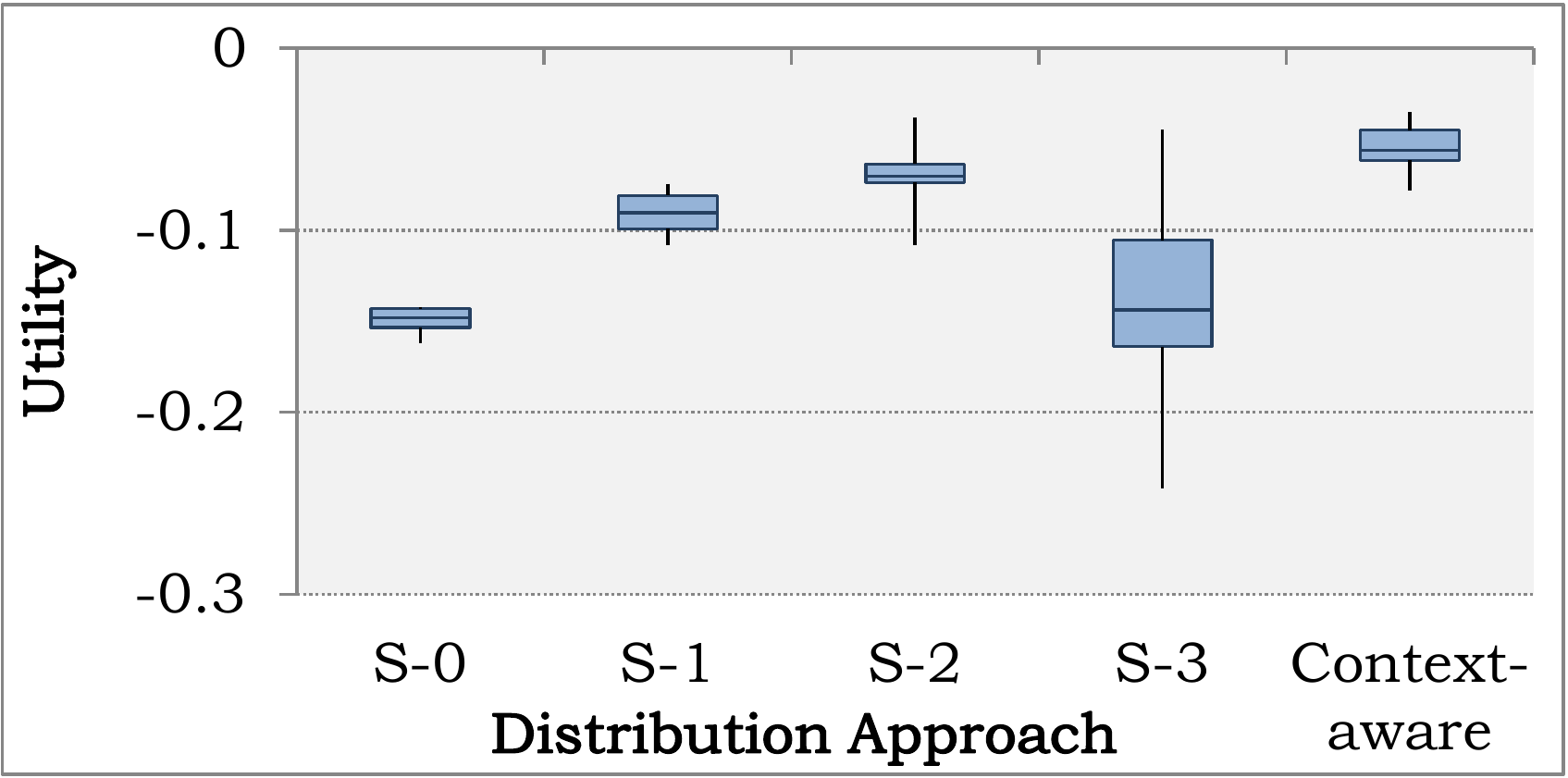}}
	\subfloat[$\lambda = 0.01$]
	{\label{fig:FDh10Lambda0.01}
	\includegraphics[width=0.33\textwidth]
	{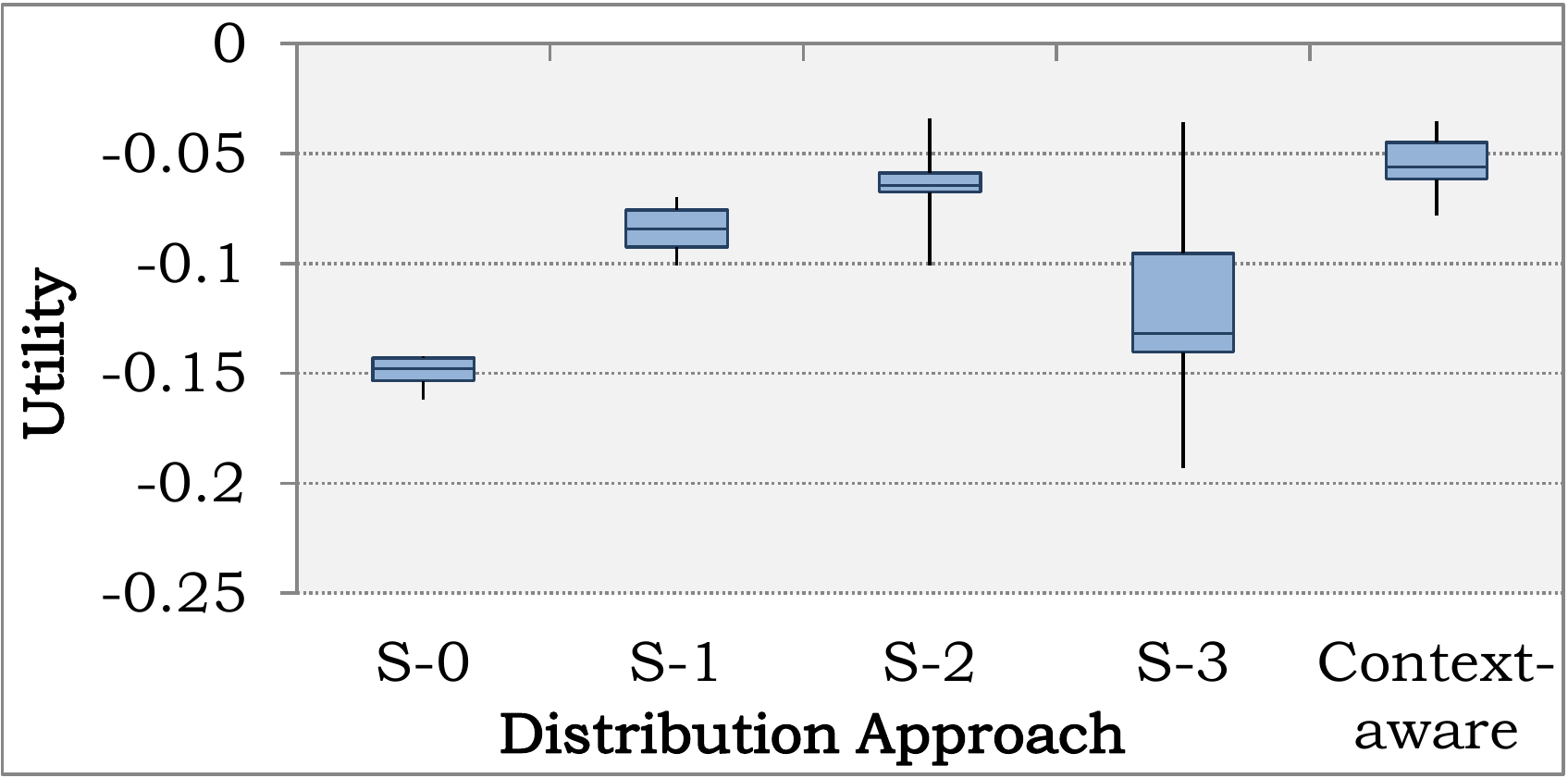}}
	\subfloat[$\lambda = 0.001$]
	{\label{fig:FDh10Lambda0.001}
	\includegraphics[width=0.33\textwidth]
	{FDh10HighLambda.pdf}}
\end{center}
\caption{Distribution of utilities over 100 experiments when $\alpha / \beta = 10$ with varying $\lambda$ for the FD application.}
\label{fig:FDhybrid10}
\end{figure*} 

\begin{figure*}
\begin{center}
	\subfloat[$\lambda = 1$]
	{\label{fig:PMh10Lambda1}
	\includegraphics[width=0.33\textwidth]
	{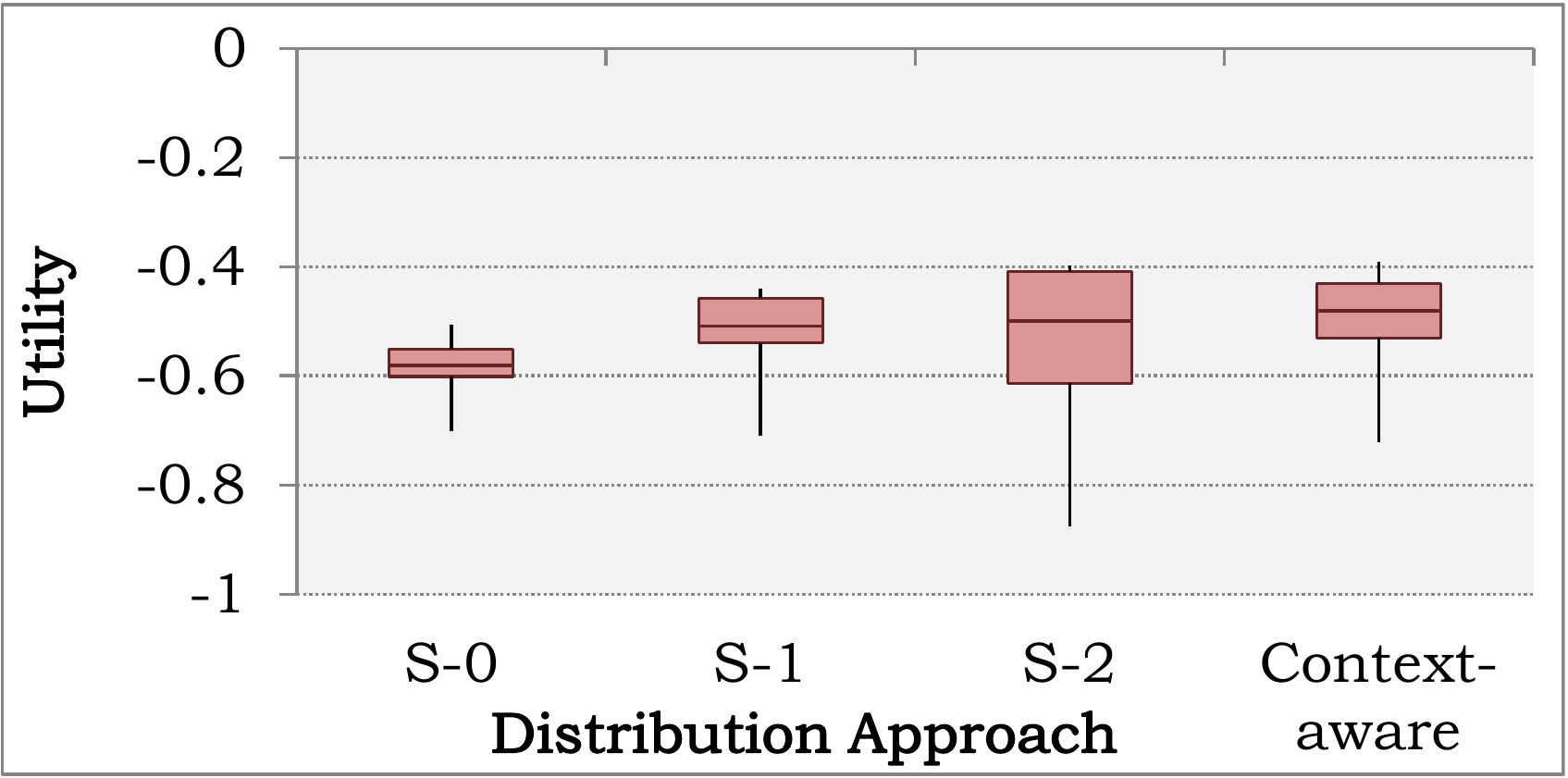}}
	\subfloat[$\lambda = 0.1$]
	{\label{fig:PMh10Lambda0.1}
	\includegraphics[width=0.33\textwidth]
	{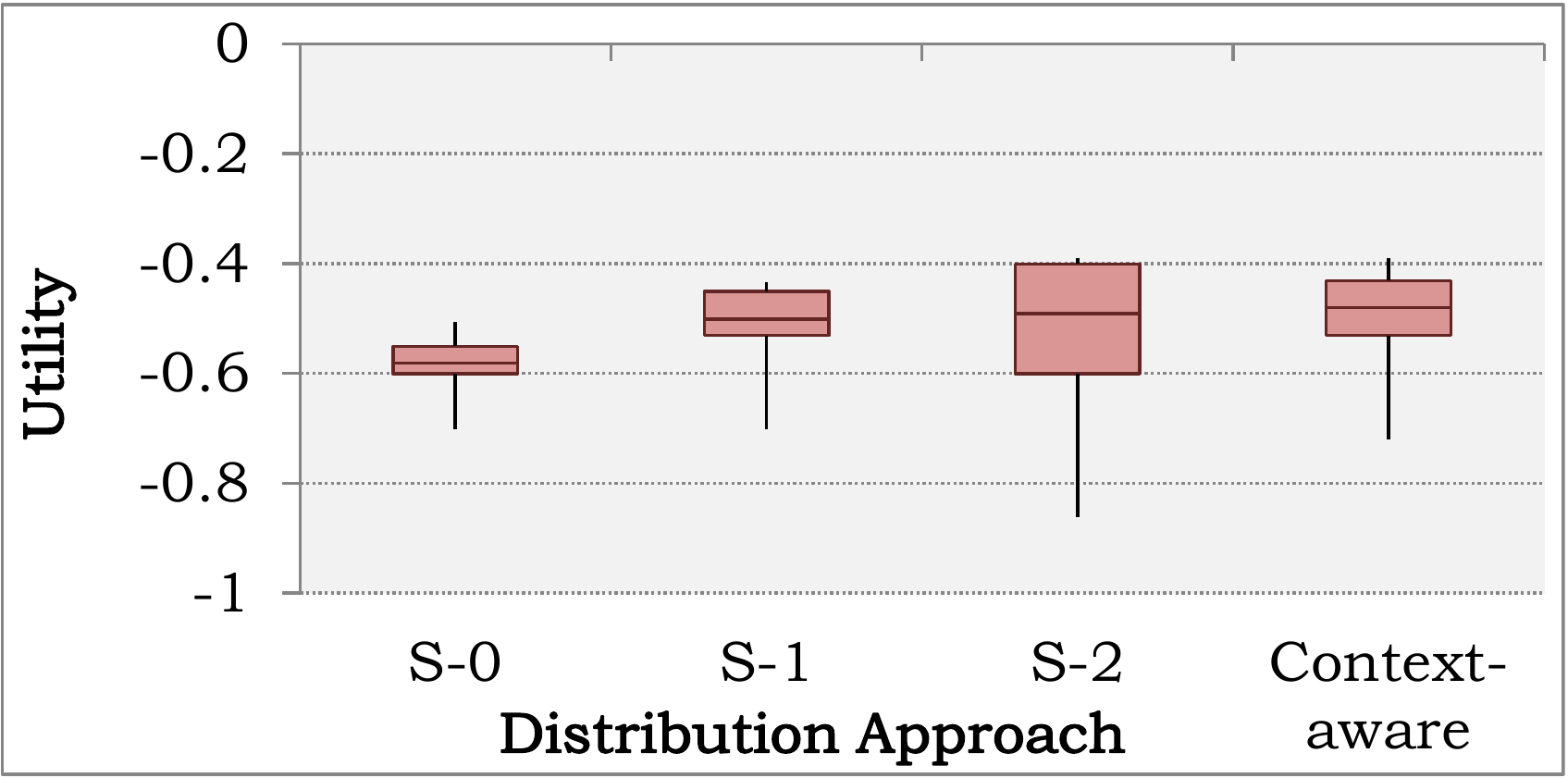}}
	\subfloat[$\lambda = 0.01$]
	{\label{fig:PMh10Lambda0.01}
	\includegraphics[width=0.33\textwidth]
	{PMh10HighLambda.pdf}}
\end{center}
\caption{Distribution of utilities over 100 experiments when $\alpha / \beta = 10$ with varying $\lambda$ for iPokeMon.}
\label{fig:PMhybrid10}
\end{figure*}

Figure~\ref{fig:FDhybrid0.1} presents the performance of FD when applying three $\lambda$ values (0.1, 0.001, 0.0001) with the hybrid strategy when $\alpha / \beta = 0.1$, i.e. when the cost is considered as 10 times important as the QoS. It is found that when $\lambda=0.1$, the gaps between the five approaches are similar to the gaps observed in the cost-aware strategy (Figure~\ref{fig:FDcostLambda0.1}). When $\lambda$ is 0.01 (Figure~\ref{fig:FDhDot1MediumLambda}) or 0.001 (Figure~\ref{fig:FDhDot1LowLambda}), the context-aware approach is no longer dominated by the most cost-efficient deployment (i.e. \textit{S-0}) and is able to benefit from the Fog-based and Fog-only deployment plans. This is because by assigning importance to the QoS factor, the context-aware approach starts to acknowledge the large reductions of application latency introduced by \textit{S-2} and \textit{S-3}. The context-aware approach, regardless of $\lambda$ values, outperforms the static approaches with the higher utility in the interquartile range.

Similarly, the iPokeMon use-case is tested with high, medium, and low $\lambda$ values (1, 0.1, 0.01) with the hybrid strategy when $\alpha / \beta = 0.1$ (Figure~\ref{fig:PMhybrid0.1}). It is found that when $\lambda=1$, the context-aware approach is dominated by \textit{S-0} (i.e. the Cloud-only deployment), as was seen in the cost-aware strategy (Figure~\ref{fig:PMcostLambda1}). When $\lambda$ is 0.1 (Figure~\ref{fig:PMhDot1MediumLambda}) or 0.01 (Figure~\ref{fig:PMhDot1LowLambda}), the application performance achieved by applying static approaches is comparable. The context-aware approach can achieve the best performance with the highest median value of the utility, though the performance gain in this use-case is less significant compared to the FD use-case. We interpret this finding as a result of the fact that the QoS improvement in iPokeMon is less significant than that in FD.

Figure~\ref{fig:FDhybrid1} displays the default hybrid strategy applied to the FD use-case, with the higher and lower values of $\lambda$ in addition to the medium value in Figure~\ref{fig:FDhybridDefault}. Since the cost and QoS factors are of equal importance in this strategy, the context-aware approach does not keep selecting the most cost-efficient \textit{S-0} as seen in the previous strategies with a high $\lambda$ value. With this hybrid strategy, regardless of $\lambda$ values, the context-aware approach can benefit from Fog-based distribution approaches. The interquartile range of the utility when applying the context-aware approach is either close to or higher than the other static approaches, which means the context-aware approach is adaptable to varying pricing methods and outperforms the static approaches. Similar finding is observed when distributing iPokeMon with the hybrid strategy with high~\ref{fig:PMh1HighLambda}, medium~\ref{fig:PMh1MediumLambda}, and low~\ref{fig:PMhybridDefault} $\lambda$ values.

When the QoS factor is considered as 10 times important as the cost factor in Equation~\ref{eq:utility}, little difference is observed when tuning $\lambda$ in both use-cases~(Figure~\ref{fig:FDhybrid10} and~\ref{fig:PMhybrid10}. The impact of $\lambda$ in this specific hybrid strategy is mitigated by the diminished importance of running cost. The benefits of the context-aware approach are again found more significant in FD than in the iPokeMon use-case. This is because the performance gain achieved by Fog-based FD is larger than the performance gain achieved by Fog-based iPokeMon.

\subsection{Summary}
The experimental results of the different distribution approaches for the chosen use-cases are summarised as follows:
\begin{itemize}[leftmargin=0.3cm]
    \item The DQN-based context-aware distribution mechanism achieves stable performance with 600+ and 400+ episodes of 20 successive deployments, for FD and iPokeMon respectively.
    \item The context-aware distribution mechanism has a small overhead (139.5ms on average), which is neglectable when compared to the job completion time measured in seconds.
    \item The context-aware distribution mechanism can effectively increase the utility when compared to all available static approaches. Its worst performance is as good as the best static approaches.
    \item When cost is the dominant factor that affects the utility, the context-aware distribution mechanism benefit from Fog-based distribution of FD only when the Fog resources are priced less than one-tenth of the Cloud resources. For the iPokeMon use-case, the advantage of context-aware distribution mechanism can be observed when the Fog resources are priced less than the Cloud resources.
    \item When QoS is the dominant factor that affects the utility, the context-aware distribution mechanism always outperforms the static approaches. The significance of the performance improved by applying the context-aware distribution mechanism depends on the performance gains achieved by utilising Fog services.
\end{itemize}

\section{Related Work}
\label{sec:relatedwork}
\begin{figure}
\begin{center}
	\includegraphics[width=0.5\textwidth]
	{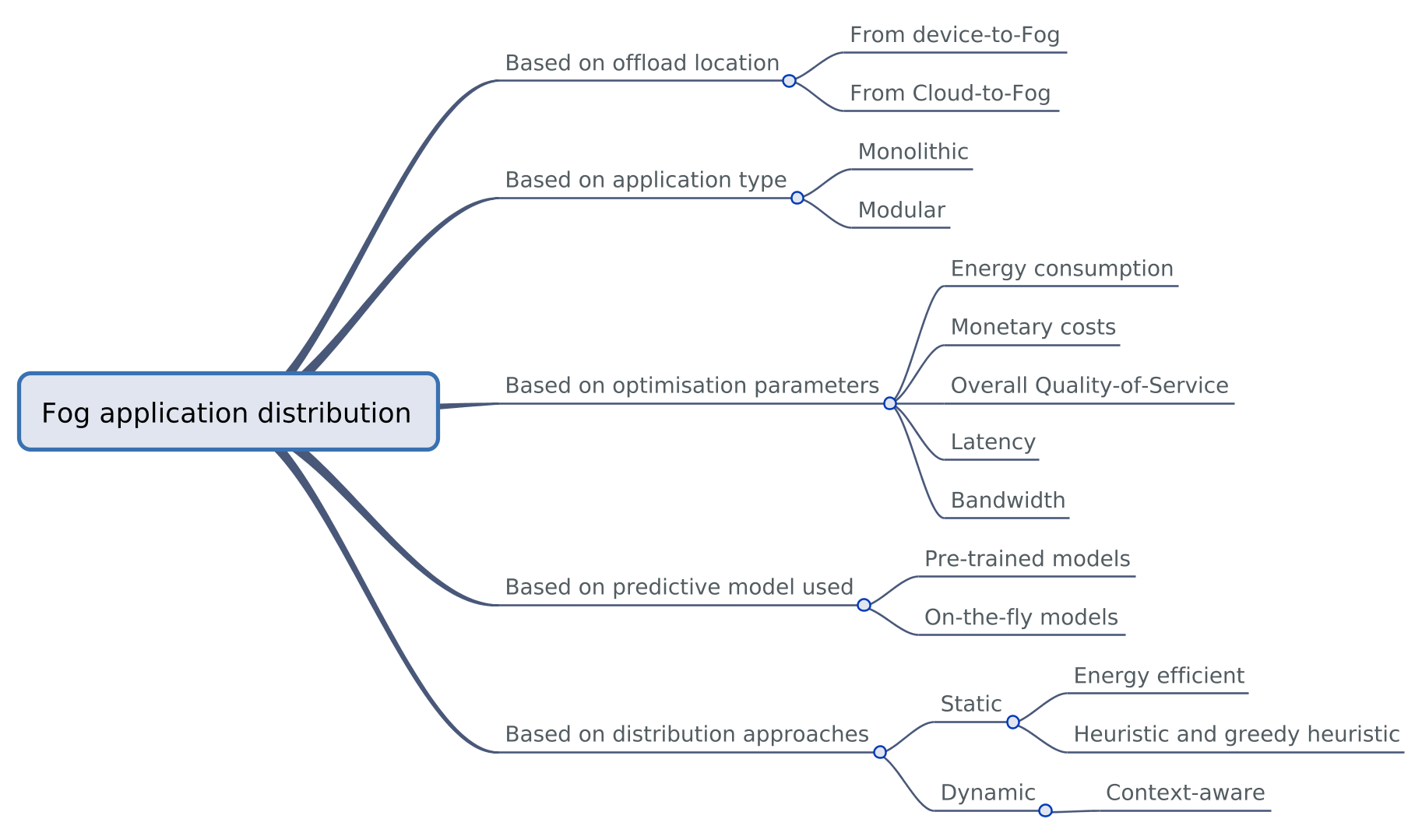}
\end{center}
\caption{A classification of existing research on Fog application distribution.}
\label{fig:relatedwork}
\end{figure}

The distribution of applications across the Cloud and the Fog (or resources located at the edge of the network) has been investigated in the context of related areas, namely Fog computing, Multi-access Edge computing (MEC) and Mobile Cloud Computing (MCC). A classification of existing research is shown in Figure~\ref{fig:relatedwork}. In this article, five classifications are highlighted based on (i) the direction of offload, (ii) the type of application, (iii) the parameters that are optimised in the problem space, (iv) the modelling techniques used, and (v) the approaches used for distributing computation workloads. 

\textit{Based on the direction of offload}, computational workloads can either be offloaded from end user devices to the MEC/Fog platform~\cite{ndikumana2019joint, wu2018noma, kumar2010cloud, chen2014decentralized} or from the Cloud to the MEC/Fog platform~\cite{lin2007enhancing, chen2015early,baguena2016towards, gao2003application}.

\textit{Based on the type of application}, either a monolithic (single application whose computational tasks cannot be distributed or divided~\cite{zhang2016energy}) or a modular (application that is composed of different services that can be geo-distributed~\cite{mahmud2018latency, hosseinpour2016approach}) application can be executed in the Fog/Edge environment. When monolithic applications are offloaded there are fewer software related challenges that need to be addressed, rather are resource challenges. For example, the application does not need to be partitioned, rather the application simply needs to be mapped to suitable resources that can meet its requirements and objectives. 

Modular applications on the other hand, require partitioning to determine feasible Fog/Edge services that can be offloaded. Such offloading strategies may use manual techniques for partitioning an application into multiple services~\cite{enorm}. For mapping the services that need to be mapped onto resources multiple techniques are proposed. For example, probing resources (Fog/Edge nodes) with micro tasks to estimate the performance when more computationally intensive services are offloaded~\cite{meurisch2017decision}; the data sizes are different for the micro tasks and the computationally intensive service. The estimations are used for a decision support mechanism. Similarly, partitioning based on task-input data is proposed to determine whether to execute the tasks locally in end devices or MEC servers~\cite{wang2016mobile}. 

\textit{Based on the parameters that are optimised} in the problem space, a variety of parameters are considered. For example, optimisation against energy consumption is well known~\cite{zhang2016energy, park2015energy}. 
Other parameters include monetary costs~\cite{do2015proximal}, the overall Quality-of-Service (QoS)~\cite{brogi2017qos},latency and bandwidth~\cite{enorm, feasibility-01}.

Another classification of Fog application distribution is \textit{based on predictive models}. 
When developing offloading strategies, it is essential to know how well the applications would perform in a given Fog computing system. Predictive models are often employed to estimate parameters, such as QoS~\cite{brogi2017qos}, cost~\cite{do2015proximal}, and energy consumption~\cite{park2015energy} of applications given the context of Fog systems. 
In order to gather training data for modelling, benchmarking is carried out on Fog systems either offline~\cite{he2018multitier, junior2019context} or online~\cite{meurisch2017decision, eom2015malmos}. A disadvantage of these methods is that they require training data to build predictive models that deal with unseen data in inference.  However, it is impractical to benchmark a Fog application on every available Fog node in a real-world setting due to the heterogeneity and the volume of Fog nodes. Therefore, the system optimisation problems are tacked by using reinforcement learning~\cite{alam2016multi, xu2017online, 1908.01275}, which provides an agent to learn on-the-fly how to behave in an environment by taking actions and seeing the results.

A model-free reinforcement learning mechanism is implemented when designing offloading policies in MEC~\cite{dinh2018distributed}. The Q-learning based algorithm in this work does not require that mobile users have prior knowledge of wireless channel information. The chosen Q-learning algorithm in this work is effective but also limited since the state variables are discretised from continuous values, which does not apply to the case when a large set of state variables is needed to account for the heterogeneity of Fog nodes. In contrast, the DQN algorithm chosen in this paper is more suitable for the complex Fog environment. 

DQN has been tested in recent works on MEC and Fog computing. A DQN-based approach to dynamically orchestrate networking, caching and computing resources for smart cities applications is employed in~\cite{he2017software}. Similarly, a DQN-based scheduling algorithm is adopted in~\cite{wang2018traffic} in order to solve the optimal offloading problem. Both of these works, however, assume that Fog applications are distributed statically and only focus on the scheduling of jobs. Our work, on the contrary, apply DQN-base solutions to a different problem of distributing Fog application across the multi-layer Fog computing environment.

\textit{Based on the approaches used for distribution}, they may either be classified as static or dynamic. Static distribution refers to when the Cloud-Fog/MEC partition of an application remains the same (i.e. the same services or modules of an application are offloaded) and does not adapt to the varying context of the Fog systems over time. Energy-efficient~\cite{zhang2016energy} and heuristic~\cite{bahreini2017efficient} techniques are considered for static deployments.  

In addition to the above static approaches, FogTorch~\cite{brogi2017qos} is designed as an offloading framework that uses a greedy heuristic approach for deriving offloading plans. This framework helps to deploy multi-component IoT applications in multi-layer Fog computing systems by searching through all possible deployment plans. However, this work is more useful at the design time when the applications are tested with all deployment plans, and it only considers network conditions as the system state. On the contrary, the context-aware distribution of Fog applications proposed in this paper works at the run time when a real-time configuration of the deployment plan is needed, and considers the state of Fog computing nodes.

\section{Conclusions}
\label{sec:conclusions}
Native Cloud applications are currently exploiting the microservices architecture in which an application is composed of multiple services that are geographically distributed. Future Cloud applications will leverage the edge of the network to improve their overall QoS in a computing model referred to as Fog computing. 

Resources located at the edge of the network, referred to as Fog nodes, are constrained in terms of capabilities (for example, number of CPU cores and memory) when compared to the Cloud and may be available intermittently. Therefore, distributing an application across the Cloud and Fog is not a trivial task given the resource limitations and the transient nature of the edge of the network. Merely an ad hoc distribution of the application could result in a poor QoS. In addition, distributing the application the same way at all times may not be efficient due to the changing environment or lack of resources to support a service on the Fog. 

The key challenge addressed in this paper is the distribution of a modular application, comprising multiple services across the Cloud and Fog in a dynamic manner. The question being addressed is `which' and `how' many services of the application should be offloaded from the Cloud to the Fog. To tackle the challenge, firstly, a mathematical model that captures the interaction between the Cloud, the Fog and end-user devices was articulated in terms of the overall QoS of the application and its running costs across the Cloud and the Fog.
Then a context-aware mechanism was proposed that dynamically generates (re)deployment plans for the application to maximise the performance efficiency of the application by taking the overall QoS and running costs into account.
The mechanism relies on deep reinforcement learning to generate a distribution strategy without prior knowledge of the available resources on the Fog nodes, network conditions, and the Fog application. 

The above context-aware distribution approach is validated on two use-cases, namely a real-time face detection application and a location-based mobile game. Both these are representative of real workload. Results obtained from an experimental evaluation highlight that the proposed context-aware distribution mechanism selects an appropriate deployment plan in near real-time and outperforms than static distribution approaches. Through the analysis of cost-efficient, QoS-aware and hybrid strategies, and the impact of Fog pricing, it is also found that the benefits of context-aware distribution mechanism are more significant when it is optimised towards QoS and/or Fog resources are priced less than Cloud resources.

\textit{Limitations and Future Work}: The limitations of the current work are: (i) Multi-tenancy is not considered in this paper. When multiple Fog applications share the same Fog node, the context changes become the result of the actions taken by all applications' Distribution Managers. In the future, we aim to extend the distribution of Fog applications as a multi-agent distribution problem that may utilise game theory and optimisation theory. (ii) The impact of different Cloud pricing models is not investigated in this paper. Other Cloud services, such as container-based services and serverless computing, maybe more cost-efficient than the VM-based services chosen in this work. Therefore, the influence of Cloud pricing models on the running costs of Fog-based deployments of applications will be explored.

\ifCLASSOPTIONcaptionsoff
  \newpage
\fi




\bibliographystyle{IEEEtran}  
\bibliography{references}

\end{document}